\documentclass[useAMS,usenatbib]{mn2e}
\pdfoutput=1
\def\kms {$\rm km\,s^{-1}$}
\def\deg {$^\circ$}

\title[Feeding vs. feedback in NGC\,1068. I. Excitation]{Feeding Versus Feedback in NGC\,1068 probed with Gemini NIFS. I. Excitation}

\author[Riffel et al.]
  {Rogemar A. Riffel$^1$, Tiberio B. Vale$^{2,3}$,
  Thaisa Storchi-Bergmann$^2$ 
\newauthor
  Peter J. McGregor$^4$\\
 $^{1}$ Universidade Federal de Santa Maria, Departamento de F\'\i sica, Centro de Ci\^encias Naturais e Exatas, 
 97105-900, Santa Maria, RS, Brazil\\
 $^{2}$ Universidade Federal do Rio Grande do Sul, Instituto de F\'\i sica, CP 15051, Porto Alegre 91501-970, RS, Brazil\\
 $^3$ Universidade Federal Fluminense, Instituto do Noroeste Fluminense de Ensino Superior, Santo Antonio de Padua, RJ - Brazil\\
 $^4$ Research School of Astronomy and Astrophysics, Australian National University, Cotter Road, Weston Creek, ACT\,2611, Australia\\
  }
\date{Released 2011}

\pagerange{\pageref{firstpage}--\pageref{lastpage}} \pubyear{2011}

\def\LaTeX{L\kern-.36em\raise.3ex\hbox{a}\kern-.15em
    T\kern-.1667em\lower.7ex\hbox{E}\kern-.125emX}

\usepackage{graphicx}

\begin{document}

\label{firstpage}

\maketitle

\begin{abstract}

We present emission-line flux distributions and ratios for the inner $\approx$\,200\,pc of the narrow-line region of the Seyfert\,2 galaxy NGC\,1068, using observations obtained with the Gemini Near-infrared Integral Field Spectrograph (NIFS) in the J, H and K bands at a spatial resolution of $\approx10\,$pc and spectral resolution of $\approx5300$. The molecular gas emission -- traced by the K-band H$_2$ emission lines -- outlines an off-centered circumnuclear ring with a radius of $\approx$\,100\,pc showing thermal excitation. The ionized gas emission lines show flux distributions mostly outlining the previously known [O\,{\sc iii}]$\lambda$5007 ionization bicone. But while the flux distributions in the H\,{\sc i} and He\,{\sc ii} emission lines are very similar to that observed in [O\,{\sc iii}], the flux distribution in the [Fe\,{\sc ii}] emission lines is more extended and broader than a cone close to the nucleus, showing a ``double bowl" or `hourglass" structure". This difference is attributed to the fact that the [Fe\,{\sc ii}] emission, besides coming from the fully ionized region, comes also from the more extended partially ionized regions, in gas excited mainly by X-rays from the active galactic nucleus. A contribution to the [Fe\,{\sc ii}] emission from shocks along the bicone axis to NE and SW of the nucleus is also supported by the enhancement of the [Fe\,{\sc ii}](1.2570\,$\mu$m)/[P\,{\sc ii}](1.1885\,$\mu$m) and [Fe\,{\sc ii}](1.2570\,$\mu$m)/Pa$\beta$ emission-line ratios at these locations and is attributed to the interaction of the radio jet with the NLR. The mass of ionized gas in the inner 200\,pc of NGC\,1068 is  $M_{\rm H\,II}\approx2.2\times10^{4}~{\rm M_\odot}$, while the mass of the H$_2$ emitting gas is only  $M_{H_2}\approx29\,{\rm M_\odot}$. Taking into account the dominant contribution of the cold molecular gas, we obtain an estimate of the total molecular gas mass of $M_{\rm cold}\approx2\,\times10^{7}\,{\rm M_\odot}$.

\end{abstract}


\section{Introduction}
\label{sec:intro}

NGC\,1068 is the brightest and most studied Seyfert\,2 galaxy, being considered the prototype of this class, after the discovery of a Seyfert\,1 spectrum in polarized light \citep{antonucci1985}, what led to the proposition of the Unified Model for  Active Galactic Nuclei (hereafter AGN) \citep{antonucci1993}. Hundreds of papers have been published on its properties, obtained from data spanning the electromagnetic spectrum from X-rays to radio wavelengths. The mass of the super-massive black hole in the nucleus of NGC\,1068 is $M_\bullet=8.6 \pm 0.6 \times 10^{6}\,{\rm M_{\odot}}$ as obtained from the keplerian motions of water maser clouds in a disk surrounding the nucleus \citep{lodato2003,kormendy11}.

Optical narrow-band images of the Narrow Line Region (NLR), obtained with ground based telescopes as well as with the Hubble Space Telescope (HST), revealed an approximately cone-shaped structure with an opening angle of $\approx 65^\circ$ and oriented along the position angle (hereafter PA) $15^\circ$ \citep[e.g.][]{pogge88,evans91,macchetto94}. The HST narrow-band [O\,{\sc iii}] $\lambda5007\AA$ image of the NLR shows several knots and filaments of enhanced emission \citep[e.g.][]{evans91,macchetto94,cecil02}. In radio wavelengths, NGC\,1068 shows a kiloparsec-scale radio jet, with the brightest structure being a compact bent nuclear radio jet, less than 1\arcsec in extent \citep[][and references therein]{gallimore96}. Although the jet leaves the nucleus at PA\,$\approx$\,15\deg, the jet is deflected at the so-called cloud C1 to PA\,$\approx$\,30\deg. \citet{gallimore2004} has shown that the nuclear source (known as component S1 in the compact radio jet) is extended and aligned perpendicularly to the radio jet. Gas kinematics obtained from long-slit HST STIS spectra reveals outflows along the [O\,{\sc iii}] bicone, with a kinematical axis aligned with the compact nuclear radio source at PA\,$\approx$\,30\deg \citep{das06,das07}. 

Optical integral field spectroscopy (IFS) of the central 10$^{\prime\prime}$ of NGC\,1068 obtained with the Gemini Multi-object Spectrograph (GMOS) at the Gemini-North telescope shows a complex flux distribution for the [O\,{\sc iii}] and H$\beta$ emitting gas, with some components associated to the bi-conical outflow, but with others attributed to high-velocity clouds and disk-like structures orbiting around the nucleus \citep{gerssen06}. 

In the near-infrared (hereafter near-IR), \citet{ms09} using  the  Spectrograph for Integral Field Observations in the Near Infrared (SINFONI) at the Very Large Telescope (VLT), at an angular resolution of 0\farcs075, found  that the molecular hydrogen H$_2$ is distributed in a ring-like structure surrounding the nucleus of the galaxy, similar to that observed in CO emission in the radio \citep[e.g.][]{schinnerer00}. In the inner 0\farcs4, \citet{ms09} found non-circular motions concluding that the H$_2$ gas streams toward the nucleus on highly elliptical or parabolic trajectories in the plane of the galaxy, estimating a mass inflow rate of $\approx$15~M$_\odot$yr$^{-1}$ \citep{ms09}.

Although the NLR of NGC\,1068 has been the subject of many studies, most of them are in the optical. We explore here the NLR properties observed in the near-IR, where we can probe distinct regions from those probed by optical observations, in particular regions with higher dust extinction. In addition, due to its proximity, we can probe the properties of the NLR down to a scale of 8\,pc with the Gemini instrument Near-infrared Integral Field Spectrograph (NIFS) used with the adaptive optics module ALTAIR. We use these observations to map the near-IR emitting gas distribution, excitation and extinction of the inner $\approx$\,200\,pc of NGC\,1068. The origin of the coronal line emission has already been discussed in \citet{mazzalay13b} using the same data presented here and we thus focus only in the properties of the low-ionization and molecular gas. The gas kinematics is discussed in \citet{barbosa14}, while the stellar population and kinematics has been presented in \citet{sb12}. Two of the strongest emission lines in the near-IR which we particularly explore are the [Fe\,{\sc ii}]$\lambda$1.644$\mu$m and H$_2\lambda2.12\mu$m, as they probe regions that are not probed in the optical. The [Fe\,II] originates in partial ionized zones, and can be present in regions where there is no H$^+$ emission; it is also sensitive to shocks and traces regions of high temperature (T$\approx$\,15000K). The H$_2$ emission probes the molecular gas, which originates in regions of much lower temperature (2000K) than those of the ionized gas. 

This paper is part of a large project in which our group AGNIFS (for Active Galactic Nuclei Integral Field
Spectroscopy)  is mapping the gas excitation and kinematics, as well as the stellar population and kinematics of the inner kiloparsec of active galaxies with IFS \citep[e.g.][]{fathi06,sb07,sb09,sb10,sb12,eso428,n4051,n7582,mrk1066pop,mrk1157b,mrk1066b,mrk1157,mrk1066a,mrk79,n5929,astor}.
We adopt a distance to NGC\,1068 of 14.4 Mpc, for which 1$^{\prime\prime}$ corresponds to 70 pc at the galaxy (for $z=0.003793$ and $H_0=75$\,km\,s$^{-1}$\,Mpc$^{-1}$).

This paper is organized as follows. In section 2 we describe the observations and data reduction.  In section 3 we present the results, which include emission-line flux and line-ratio maps for the near-IR emission lines.  In section 3  we derive physical parameters for the NLR and discuss the origin of the [Fe{\sc\,ii}] and H$_2$ emission. The conclusions of this work are presented in Section 5.

\section{Observations and Data Reduction}
\label{sec:observ}

The spectroscopic data consists of a number of data cubes in the near-IR bands, obtained at the Gemini North telescope, with the Near Infrared Integral Field Spectrometer (hereafter NIFS), described in \cite{nifs03}, operating with the ALTAIR adaptive optics module. NIFS has a square field of view of $\approx$ 3\arcsec\ $\times$ 3\arcsec, divided into 29 slitlets, each of which is sampled by 69 detector pixels (along each slitlet). Each single slitlet is 0.103\arcsec\ wide and each detecting pixel measures 0.044\arcsec\ (spatial sampling). ALTAIR was used in its Natural Guide Star mode and the wave front sensor was fed with optical light from the nucleus of NGC\,1068. 

\begin{figure}
    \centering
    \includegraphics[scale=0.57]{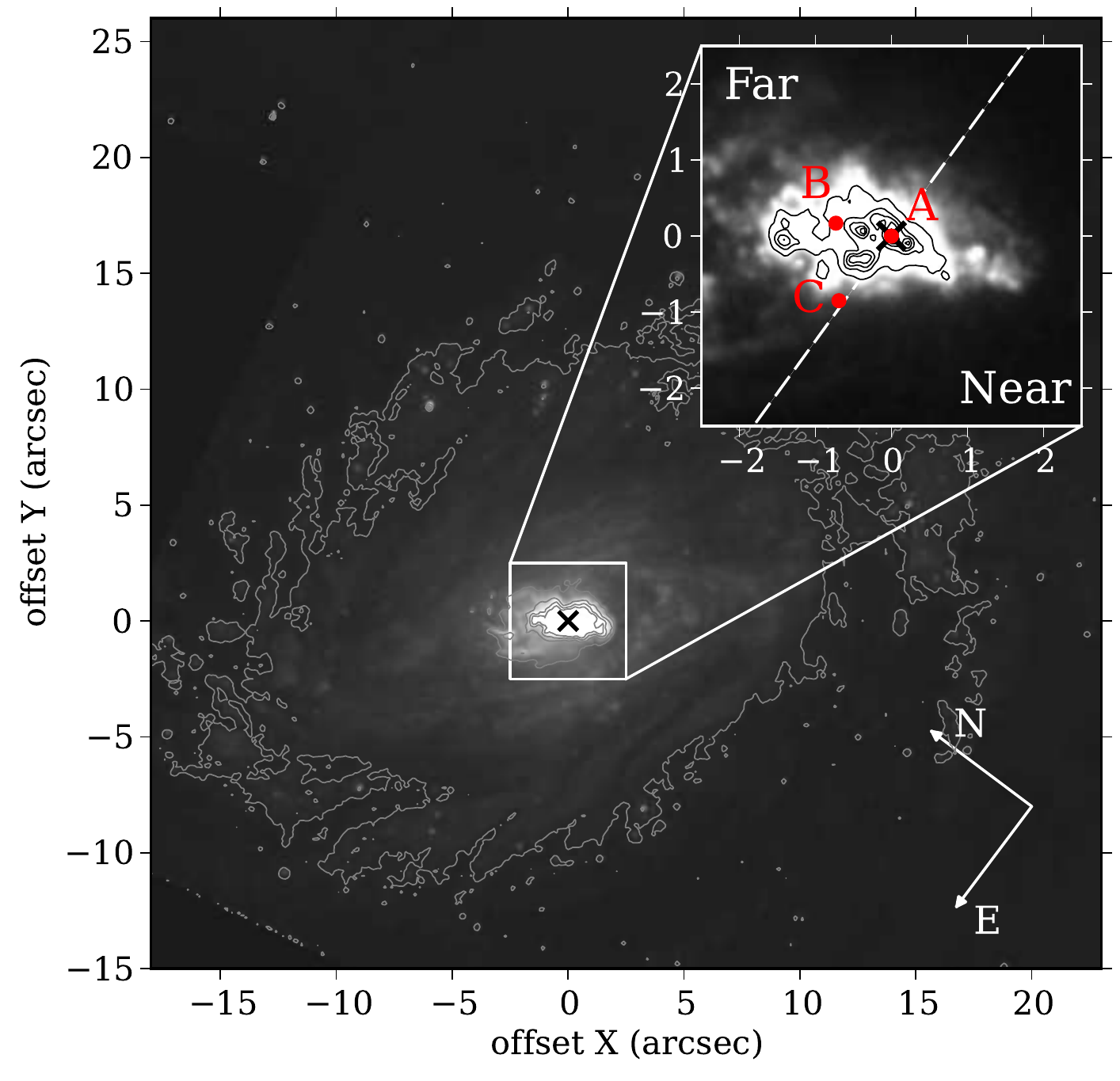}
    \caption{The large panel shows an HST F606W image of NGC\,1068 with the field-of-view of our observations indicated by the square. The insert is the [O{\sc\,iii}]\,5007\AA\ flux distribution, from \citet{sda2003}, with 6 linearly spaced contours overplotted. The three red dots indicate the positions from which we obtained the spectra shown in Fig.\,\ref{fig:sample_spectra}. The black cross shows the position of the galaxy nucleus. The white dashed line shows the orientation of the major axis of the galaxy, with near and far sides indicated.}
    \label{fig:nuclearzoom}
\end{figure}

The data were obtained on the nights of November 26, December 03 and 09, 2008, and covered the standard J, H, K spectral bands at two-pixel resolving powers of 6040 for the J band and 5290 for the other bands. This resulted in an effective wavelength coverage of 1.15-1.34 $\mu$m, 1.48-1.79 $\mu$m, 1.93-2.34 $\mu$m, respectively. Additional spectra were obtained at the K$_{long}$ setting of the K grating. This covers the wavelength range 2.11-2.52 $\mu$m, which includes the  Q-branch H$_2$ emission lines.

Dithering was used in the observations in order to cover a field-of-view (hereafter FOV) of $\approx$ 5\arcsec\ $\times$ 5\arcsec. Fig. \ref{fig:nuclearzoom} shows this field-of-view on top of an  HST F606W image of NGC\,1068 obtained with the Planetary Camera. The insert shows the intensity distribution in the [O{\sc\,iii}]$\lambda$5007 emission line, from \citet{sda2003}, with 6 contours overplotted, within the FOV of our observations. The white dotted line shows the orientation of the photometric major axis of the outer disk, at PA$=80$\deg$\pm$5\deg \citep{emsellem06}, which is also consistent with the stellar kinematic major-axis within our FOV \citep{sb12}, with the near and far sides of the galaxy disk indicated. The final spatial coverage of the data cubes were slightly different in each spectral band depending on the dithering process: 49 $\times$ 118 pixels in the J band, 50 $\times$ 117 pixels in the H band, K$_s$ was 49 $\times$ 118 pixels in the K$_s$ band and 50 $\times$ 118 pixels in the K$_l$ band.

The instrument was set to a position angle of 300$^\circ$, resulting in a finer spatial sampling perpendicular to the ionization cone.
The full width at half maximum (FWHM) of the spatial profile of the telluric standard star is $0\farcs11\,\pm\,0\farcs02$ in the H and K bands, corresponding to $\approx 8\,$pc at the galaxy, while in the J band it is larger, $0\farcs14\,\pm\,0\farcs02$, corresponding to $\approx 10\,$pc at the galaxy.

The data reduction followed standard procedures and was accomplished using tasks from the {\sc nifs} package -- which is part of the {\sc gemini iraf} package, as well as generic {\sc iraf} tasks. The reduction included trimming of the images, flat-fielding, sky subtraction, wavelength and s-distortion calibrations, and remotion of the telluric absorptions. The flux calibration was done by interpolating a black body function to the spectrum of the telluric standard star. The individual data cubes in each band were then median combined to a single data cube. More details on the data reduction process can be found in \citet{sb12}. The final data cubes contain $\approx$ 5800 spectra, with each spectrum corresponding to $0\farcs103 \times 0\farcs044$ in the sky, and  8$\times$3.4 pc$^2$ at the galaxy. The entire field of $\approx$ 5\arcsec\ $\times$ 5\arcsec covered by the observations corresponds to a region of $\approx$ 400\,pc $\times$ 400\,pc at the galaxy.

\section{Results}
\label{sec:results}

\begin{figure*}
    \includegraphics[scale=0.6]{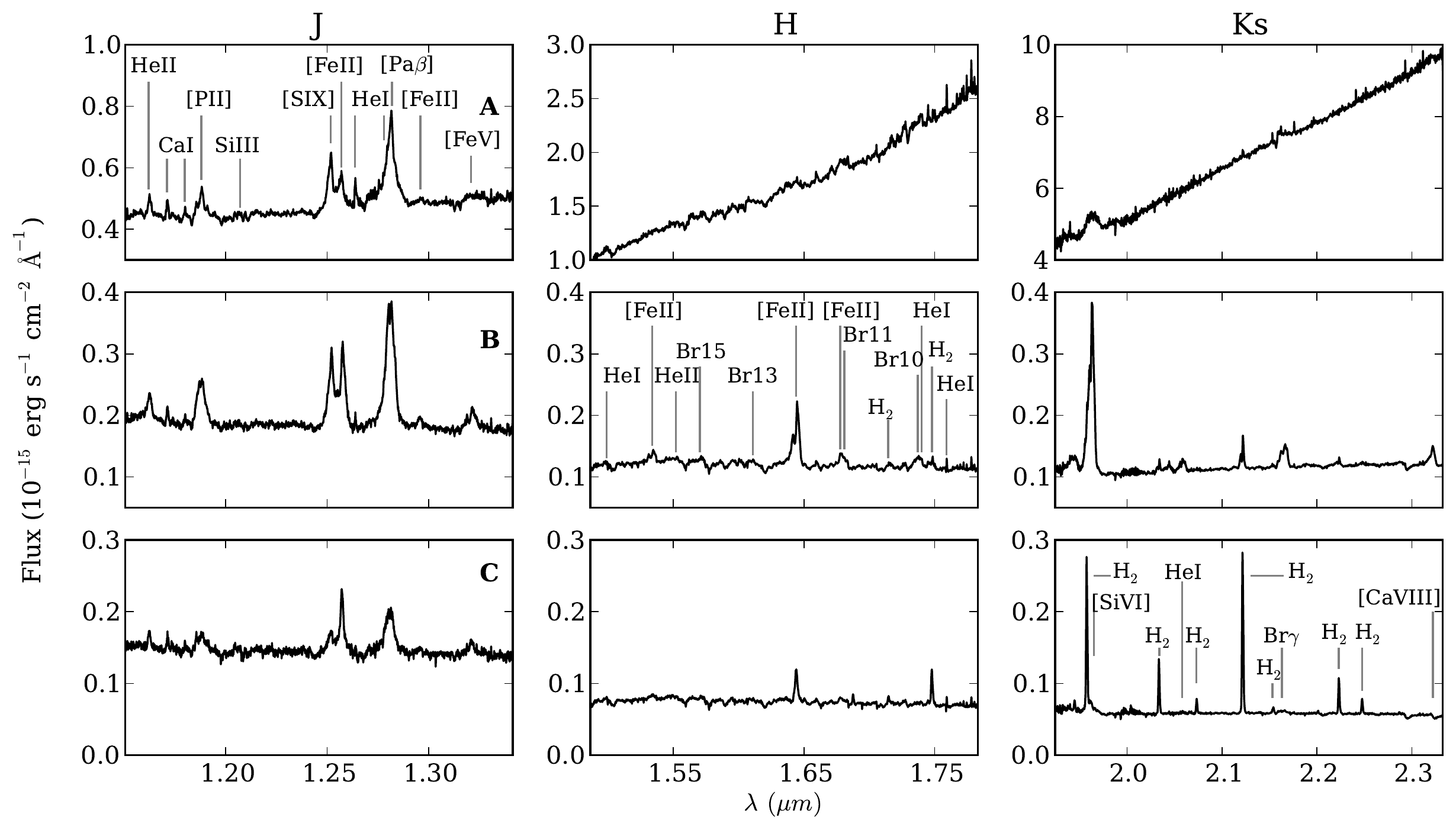}
    \caption{Sample of integrated spectra within apertures of 0.3\arcsec $\times$ 0.3\arcsec\ from three positions: (A-top) the nucleus; (B) the position of the peak flux in the [Fe{\sc\,ii}] $1.257 \mu$m emission line; and (C) the position of the peak flux of the H$_2\lambda2.1218 \mu$m emission line. These positions are identified in Fig.\,\ref{fig:nuclearzoom}.}
    \label{fig:sample_spectra}
\end{figure*}

In Fig. \ref{fig:sample_spectra} we present sample spectra extracted from $0\farcs3\,\times\,0\farcs3$ apertures, centered at the three positions shown in Fig.~\ref{fig:nuclearzoom}: (A) the continuum peak, which was adopted as the position of the nucleus; (B) the position where [Fe{\sc\,ii}] $\lambda 1.257\mu$m emission line reaches its maximum intensity at 0\farcs55 north-east from the nucleus; and (C) the position where the H$_2$ $\lambda 2.1218\mu$m emission line reaches its maximum intensity at 1\farcs2 south-east from the nucleus. The main emission lines seen in the J, H and K bands are identified and labeled in the spectra. 
The strongest emission lines of the J band are identified in the top-left panel, while in the middle central panel we identify the strongest H-band emission lines and in the bottom-right panel the strongest K-band lines. The top central and right panels show the strong red continuum present in the H and K band nuclear spectra, attributed to the nuclear (torus) dust component, discussed in \citet{sb12}. In the extra-nuclear spectra shown in the middle and bottom panels, the continuum is much flatter and even blue in the case of the position C, where we have found the presence of young stars in \citet{sb12}.

In Table \ref{tab:fluxes} we list the flux of the emission lines we could measure in the integrated spectra from positions A, B and C in Fig.~\ref{fig:nuclearzoom}. They were obtained by integrating the flux under each emission line profile after the subtraction of the underlying continuum contribution, which was determined by a linear fit (least-squares fitting) to the observed continuum in spectral windows adjacent to each emission line. In the case of blended lines, we defined a common adjacent continuum and fitted multiple gaussians. 
The fluxes are presented in units of 10$^{-15}$ erg\,cm$^{-2}$\,s$^{-1}$ and the listed errors are due to  the uncertainty associated to the fitting procedure.

\begin{table*}
\begin{scriptsize}
\centering
\begin{tabular}{l l l r r r r r r}
\hline
   $\lambda_{\mathrm{vac}}$  ($\mu$m)  & {\bf Ion}   &    {\bf Line ID ($J_i-J_k$)}   &    (A) Nucleus  & (B) [Fe\,{\sc ii}] peak & (C) H$_2$ peak\\
  \hline
  1.1630 &             He{\sc\,ii} &                       $5-7$ &     2.14  $\pm$   0.19 &     2.03  $\pm$   0.10 &     0.706  $\pm$   0.10 \\
  1.1714 &              Ca{\sc\,i} &              $^3D_1-^3Po_1$ &     1.22  $\pm$   0.25 &     0.46  $\pm$   0.12 &     0.423  $\pm$   0.12 \\
  1.1799 &              Ca{\sc\,i} &              $^3Po_2-^3D_2$ &     0.79  $\pm$   0.23 &     0.35  $\pm$   0.12 &     0.187  $\pm$   0.10 \\
  1.1855 &            He{\sc\,ii}  &                      $7-29$ &     1.81  $\pm$   0.29 &     --                 &     0.513  $\pm$   0.12 \\
  1.1886 &           [P{\sc\,ii}]  &               $^3P_2-^1D_2$ &     4.83  $\pm$   0.29 &     8.25 $\pm$   2.50  &     1.988  $\pm$   0.15 \\
  1.1909 &            Fe{\sc\,ii}  &   $e^6G_{13/2}-^6Fo_{11/2}$ &     1.62  $\pm$   0.29 &     --                 &     0.298  $\pm$   0.12 \\
  1.2068 &            Si{\sc\,iii} &              $^1D_2-^1Po_1$ &     0.33  $\pm$  0.15  &     0.55  $\pm$   0.14 &     0.629  $\pm$   0.14 \\
  1.2147 &             Si{\sc\,iv} &  $^2Po_{3/2}-^2D_{3/2-5/2}$ &     --                 &     0.74  $\pm$   0.10 &     0.682  $\pm$   0.10 \\
  1.2523 &            [S{\sc\,ix}] &               $^3P_2-^3P_1$ &     11.9 $\pm$   2.1   &    10.34  $\pm$   0.50 &     3.565  $\pm$   0.26 \\
  1.2570 &           [Fe{\sc\,ii}] &     $a^6D_{9/2}-a^4D_{7/2}$ &     8.47 $\pm$   1.82  &     8.62  $\pm$   0.46 &     4.465  $\pm$   0.35 \\
  1.2640 &              He{\sc\,i} &          $^3S_1-^3Po_{0-2}$ &     1.53  $\pm$   0.50 &    --                  &     0.143  $\pm$   0.16 \\
1.2822 & H{\sc\,i}\,Pa$\beta^*$ &                          $3-5$ &  30.6 $\pm$   5.3     &   25.80$\pm$ 8.20      &          7.26  $\pm$  0.52 \\
  1.2964 &           [Fe{\sc\,ii}] &     $b^2P_{3/2}-c^2G_{7/2}$ &     --                 &     1.22  $\pm$   0.12 &     0.724  $\pm$   0.10 \\
  1.3216 &            [Fe{\sc\,v}] &               $^5P_2-^3P_0$ &     --                 &     3.98  $\pm$   0.18 &     1.811  $\pm$   0.14 \\
  1.5330 &          [Fe{\sc\,ii}]  &     $a^4F_{9/2}-a^4D_{5/2}$ &     --                 &     1.78  $\pm$   0.06 &     0.64  $\pm$   0.06 \\
  1.5492 &             He{\sc\,ii} &                      $8-33$ &     --                 &     1.05  $\pm$   0.08 &     0.86  $\pm$   0.07 \\
  1.5705 &         H{\sc\,i}\,Br15 &                      $4-15$ &     --                 &     0.48  $\pm$   0.06 &     0.19  $\pm$   0.04 \\
  1.5953 &              He{\sc\,i} &          $^3S_1-^3Po_{0-2}$ &     --                 &     0.55  $\pm$   0.07 &     0.34  $\pm$   0.04 \\
  1.6028 &            Fe{\sc\,ii}] &    $v^4Fo_{5/2}-e^4H_{7/2}$ &     --                 &     0.38  $\pm$   0.06 &     0.14  $\pm$   0.04 \\
  1.6085 &            He{\sc\,i}   &          $^3D_3-^3Fo_{2-4}$ &     --                 &     0.41  $\pm$   0.08 &     0.26  $\pm$   0.06 \\
  1.6114 &       H{\sc\,i}\,Br13   &                      $4-13$ &     1.00  $\pm$   0.25 &     0.41  $\pm$   0.08 &     0.12  $\pm$   0.07 \\
  1.6388 &         [Fe{\sc\,ii}]   &     $b^4P_{5/2}-b^2P_{1/2}$ &     --                 &     7.31  $\pm$   0.20 &     1.31  $\pm$   0.12 \\
  1.6440 &         [Fe{\sc\,ii}]   &     $a^4F_{9/2}-a^4D_{7/2}$ &     --                 &     10.17 $\pm$   1.25 &     2.44  $\pm$   0.15 \\
  1.6596 &             He{\sc\,ii} &                      $8-23$ &     2.59  $\pm$   0.37 &     0.47  $\pm$   0.06 &     0.33  $\pm$   0.06 \\
  1.6713 &           [Fe{\sc\,iv}] &       $^4D_{7/2}-^2D_{3/2}$ &     3.04  $\pm$   0.80 &     --                 &     --                  \\
  1.6773 &         [Fe{\sc\,ii}]   &     $a^4F_{7/2}-a^4D_{5/2}$ &     --                 &     2.85  $\pm$   0.10 &     0.75  $\pm$   0.06 \\
  1.6811 &       H{\sc\,i}\,Br11   &                      $4-11$ &     --                 &     2.75  $\pm$   0.50 &     0.37  $\pm$   0.06 \\
  1.6877 &                   H$_2$ &                 $1-0\,S(9)$ &     --                 &     0.06  $\pm$   0.04 &     0.34  $\pm$   0.04 \\
  1.6918 &              He{\sc\,i} &              $^1Po_1-^1D_2$ &     --                 &     0.33  $\pm$   0.04 &     0.18  $\pm$   0.07 \\
  1.6985 &            He{\sc\,i}   &               $^1Po_1-1S_0$ &     --                 &     0.59  $\pm$   0.08 &     0.10  $\pm$   0.06 \\
  1.7058 &           He{\sc\,ii}   &                      $8-21$ &     --                 &     0.13  $\pm$   0.08 &     0.24  $\pm$   0.06 \\
  1.7147 &                   H$_2$ &                 $1-0\,S(8)$ &     0.23  $\pm$   0.15 &     0.22  $\pm$   0.04 &     0.29  $\pm$   0.04 \\
  1.7334 &              He{\sc\,i} &          $^3D_3-^3Fo_{2-4}$ &     --                 &     4.33 $\pm$   1.25  &     0.34  $\pm$   0.04 \\
  1.7360 &           He{\sc\,ii}   &                      $8-20$ &     --                 &     0.18  $\pm$   0.06 &     0.21  $\pm$   0.04 \\
  1.7367 &       H{\sc\,i}\,Br10   &                      $4-10$ &     --                 &     1.99  $\pm$   0.06 &     0.17  $\pm$   0.04 \\
  1.7454 &            He{\sc\,i}   &              $^3S_1-^3Po_0$ &     --                 &     0.41  $\pm$   0.08 &     0.21  $\pm$   0.10 \\
  1.7480 &                 H$_2$   &                 $1-0\,S(7)$ &     1.09  $\pm$   0.76 &     0.61  $\pm$   0.08 &     1.55  $\pm$   0.10 \\
  1.9439 &              He{\sc\,i} &              $^1Fo_3-^1G_4$ &     --                 &     3.78  $\pm$   0.16 &     0.92  $\pm$   0.12 \\
  1.9548 &            He{\sc\,i}   &              $^3D_1-^3Po_0$ &     --                 &     1.96  $\pm$   0.42 &     --  \\
  1.9576 &                 H$_2$   &                 $1-0\,S(3)$ &     --                 &     7.71  $\pm$   0.53 &     5.67  $\pm$   0.18 \\
  1.9598 &            He{\sc\,i}   &              $^1Po_1-^1D_2$ &     --                 &     3.10  $\pm$   0.49 &     2.09  $\pm$   0.15 \\
  1.9650 &         [Si{\sc\,vi}]   &     $^2Po_{3/2}-^2Po_{1/2}$ &     --                 &    27.66  $\pm$   1.23 &     1.40  $\pm$   0.14 \\
  2.0338 &                 H$_2$   &                 $1-0\,S(2)$ &     --                 &     0.72  $\pm$   0.11 &     2.30  $\pm$   0.22 \\
  2.0444 &            He{\sc\,i}   &          $^3S_1-^3Po_{0-2}$ &     --                 &     1.13  $\pm$   0.10 &     0.24  $\pm$   0.06 \\
  2.0587 &              He{\sc\,i} &              $^1S_0-^1Po_1$ &     --                 &     2.34  $\pm$   0.10 &     0.59  $\pm$   0.06 \\
  2.0735 &                   H$_2$ &                 $2-1\,S(3)$ &     --                 &     0.33  $\pm$   0.06 &     0.63  $\pm$   0.04 \\
  2.1204 &           He{\sc\,ii}   &                      $9-25$ &     --                 &     1.02  $\pm$   0.12 &     --                  \\
  2.1218 &                 H$_2$   &                 $1-0\,S(1)$ &     2.81  $\pm$   0.17 &     2.13  $\pm$   0.12 &     6.95  $\pm$   0.22 \\
  2.1542 &                   H$_2$ &                 $2-1\,S(2)$ &     --                 &     0.34  $\pm$   0.08 &     0.37  $\pm$   0.05 \\
  2.1661 & H{\sc\,i}\,Br$\gamma^*$   &                       $4-7$ &     12.81  $\pm$4.30 &     8.82 $\pm$   1.50  &     0.69  $\pm$   0.04 \\
  2.1766 &              Li{\sc\,i} &   $^2D_{5/2}-2Po_{1/2-3/2}$ &     --                 &     0.57  $\pm$   0.06 &     0.09  $\pm$   0.03 \\
  2.1891 &             He{\sc\,ii} &                      $7-10$ &     --                 &     0.57  $\pm$   0.04 &     --                 \\
  2.2233 &                 H$_2$   &                 $1-0\,S(0)$ &     --                 &     1.24  $\pm$   0.05 &     1.79  $\pm$   0.04 \\
  2.2477 &                  H$_2$  &                 $2-1\,S(1)$ &     --                 &     0.45  $\pm$   0.06 &     0.73  $\pm$   0.07 \\
  2.3163 &            He{\sc\,i}   &          $^3D_3-^3Po_{0-2}$ &     --                 &     1.18  $\pm$   0.05 &     0.17  $\pm$   0.06 \\
  2.3210 &       [Ca{\sc\,viii}]   &     $^2Po_{1/2}-^2Po_{3/2}$ &     --                 &     7.55 $\pm$   2.50  &     --                  \\
  2.4066 &                   H$_2$ &                 $1-0\,Q(1)$ &     2.34  $\pm$   0.18 &     2.86  $\pm$   0.26 &     5.80  $\pm$   0.22 \\
  2.4134 &                   H$_2$ &                 $1-0\,Q(2)$ &     2.03  $\pm$   0.32 &     0.56  $\pm$   0.26 &     1.84  $\pm$   0.16 \\
  2.4237 &                   H$_2$ &                 $1-0\,Q(3)$ &     1.36  $\pm$   0.24 &     2.70  $\pm$   0.14 &     5.99  $\pm$   0.06 \\
  2.4375 &                   H$_2$ &                 $1-0\,Q(4)$ &     --                 &     0.41  $\pm$   0.08 &     1.59  $\pm$   0.08 \\
  2.4548 &                   H$_2$ &                 $1-0\,Q(5)$ &     1.22  $\pm$   0.40 &     1.74  $\pm$   0.10 &     4.14  $\pm$   0.26 \\
  2.4755 &                 H$_2$   &                 $1-0\,Q(6)$ &     --                 &     1.17  $\pm$   0.38 &     0.97  $\pm$   0.12 \\
  2.4810 &           He{\sc\,ii}   &                     $10-35$ &     --                 &    19.34  $\pm$   1.25 &     2.95  $\pm$   0.15 \\
  2.4817 &            He{\sc\,i}   &              $^1Po_1-^1D_2$ &     --                 &     6.75  $\pm$   0.39 &     0.35  $\pm$   0.12 \\
  2.4833 &        [Si{\sc\,vii}]   &               $^3P_2-^3P_1$ &     1.20  $\pm$   0.45 &    32.41  $\pm$   1.39 &     0.20  $\pm$   0.12 \\
  \hline
\multicolumn{7}{l}{*: Pa$\beta$ and Br$\gamma$ lines seem to have a broad component, although narrower than that seen in the optical }\\
\multicolumn{7}{l}{polarized spectra, and may just be the result of the blend of several components.}\\
\end{tabular}                                       
\end{scriptsize} 
\caption{Fluxes of emission lines in the J, H, K$_s$ and K$_l$ bands measured within 0$\farcs3 \times 0\farcs 3$ apertures at three positions: (A) the nucleus, (B) the location of the  [Fe\,{\sc ii}] emission peak  at 0\farcs55 north-east from the nucleus, and (C) the location of the H$_2$ emission peak at 1\farcs2 south-east from the nucleus, as shown in Fig.\,\ref{fig:nuclearzoom}. Flux units are 10$^{-15}$ erg\,cm$^{-2}$\,s$^{-1}$. 
The Line ID column shows the spectroscopic term for the lower level $J_i$ and upper level $J_k$. Flux values were obtained via integration of Gaussian profiles fitted to the lines, after subtraction of the continuum contribution. In the case of blended/multicomponent lines, multiple gaussians were fitted.}
\label{tab:fluxes}
\end{table*}

\subsection{Emission-line flux distributions}
\label{subsec:emission}

\begin{figure*}
  \begin{tabular}{cc}
    \includegraphics[scale=0.47]{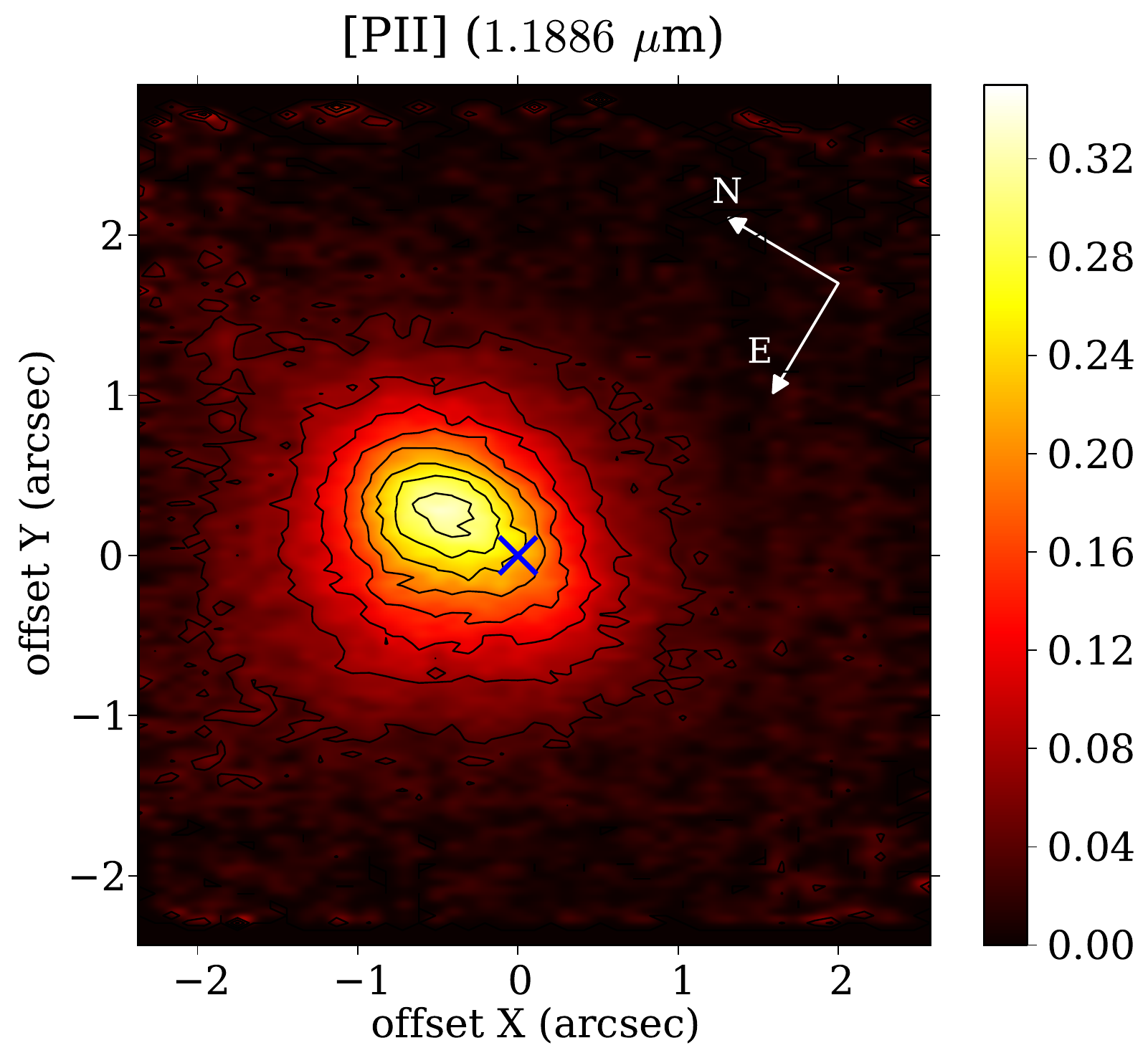}&
    \includegraphics[scale=0.47]{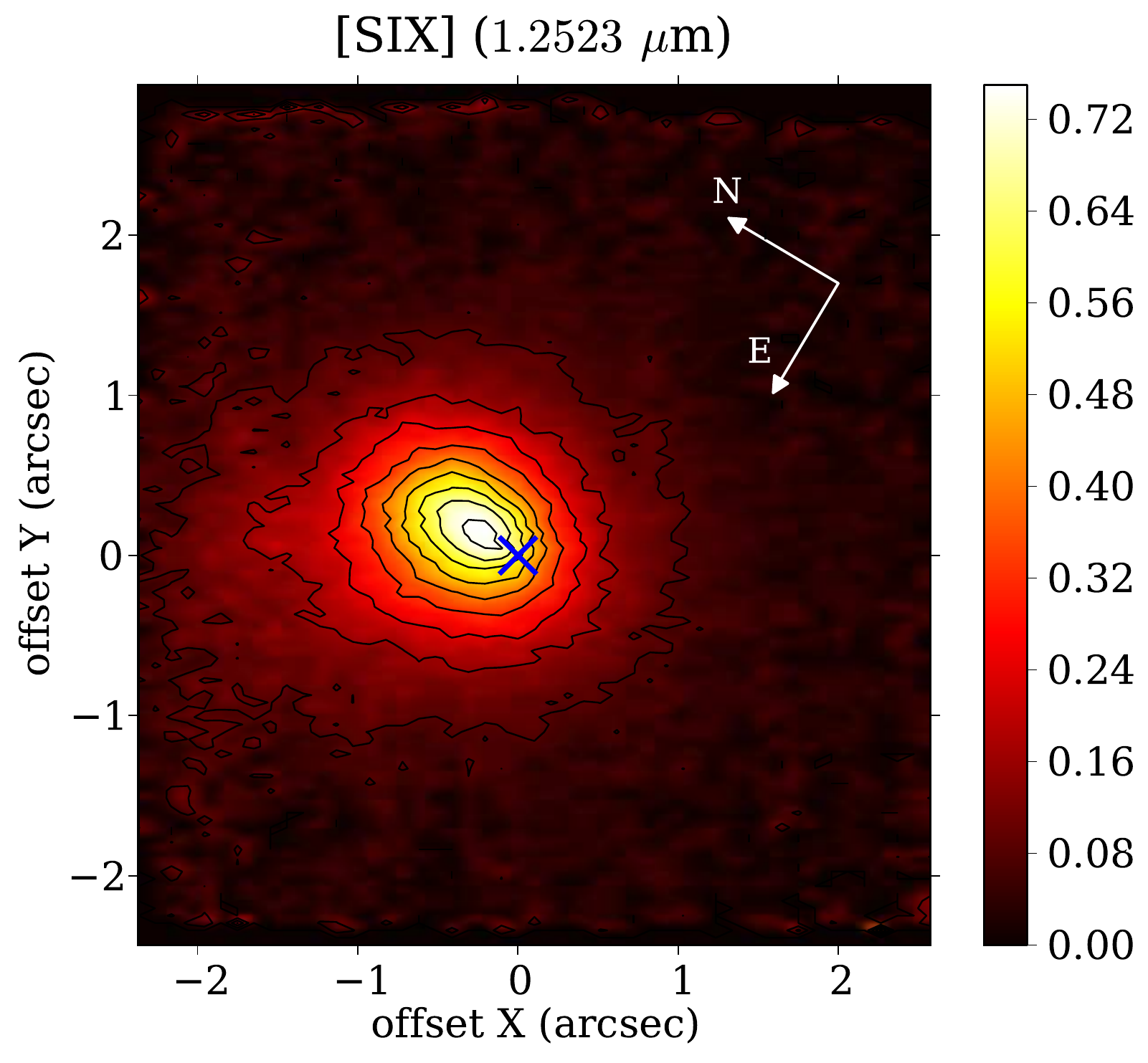}\\
    \includegraphics[scale=0.47]{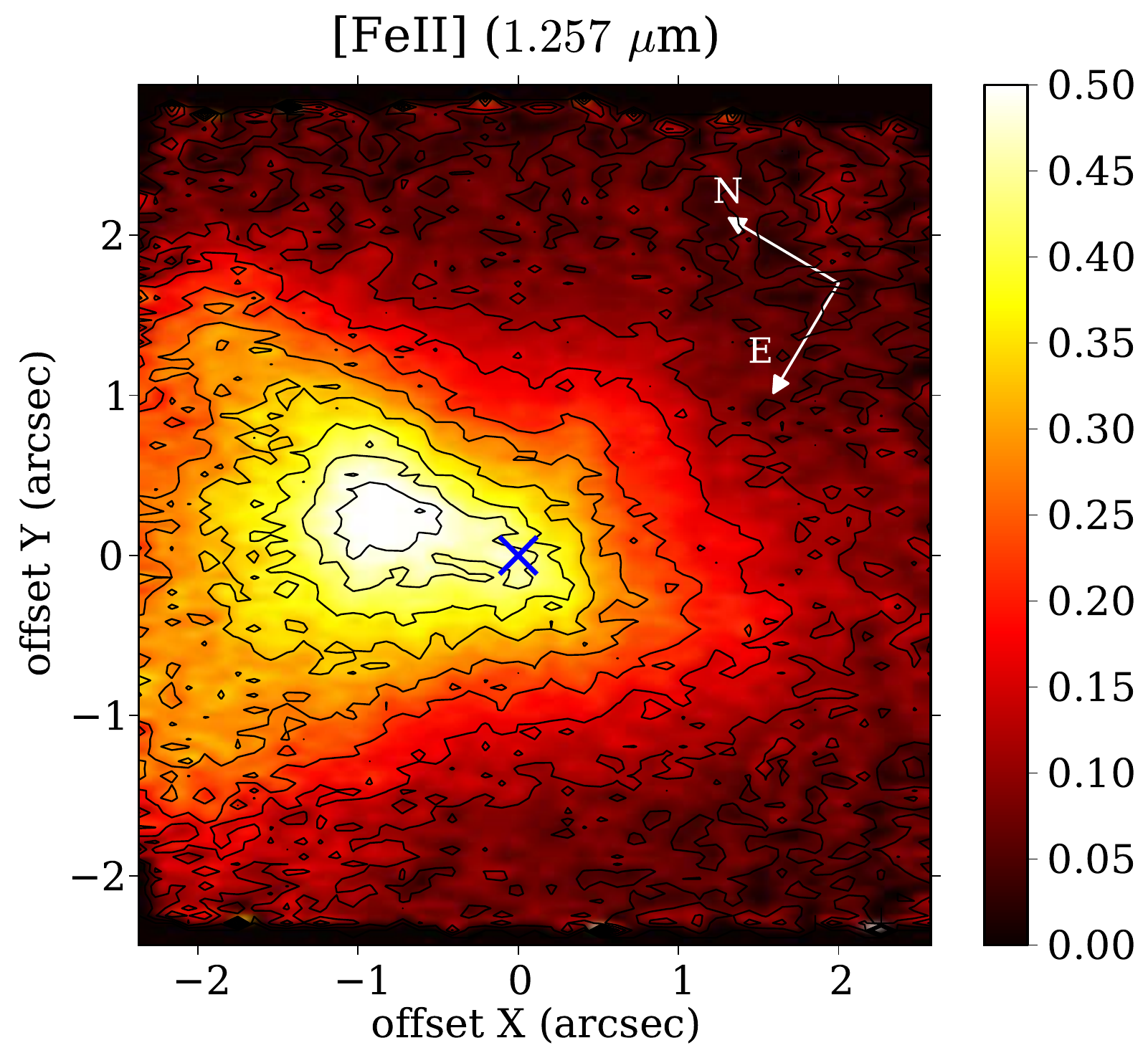} &
    \includegraphics[scale=0.47]{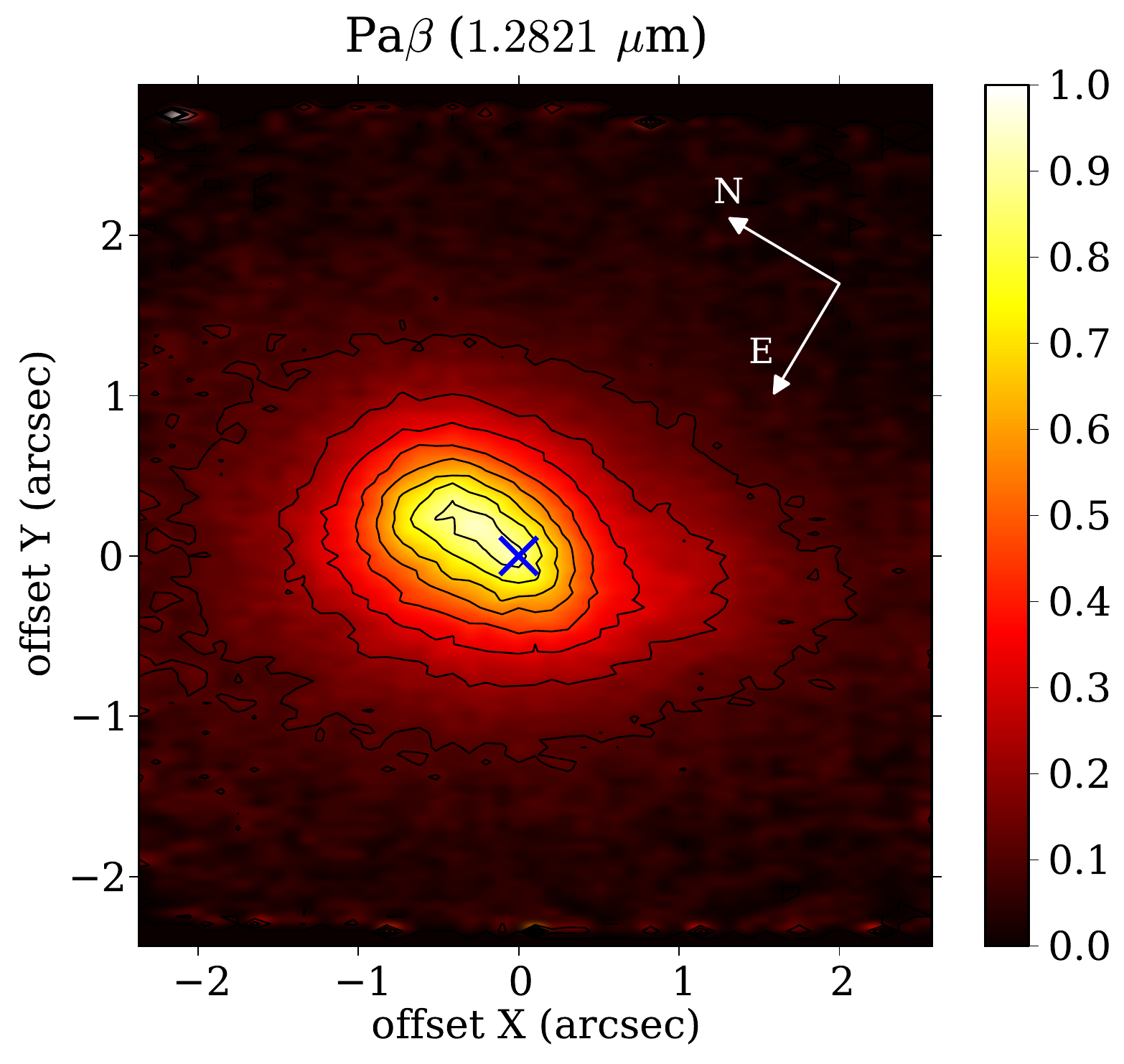} 
  \end{tabular}
  \caption{Flux distributions in the J-band emission lines: [P{\sc\,ii}] ($1.1886\mu$m), [S{\sc\,ix}] ($1.2523\mu$m), [Fe{\sc\,ii}] ($1.257\mu$m) and  Pa$\beta$($1.2821\mu$m)
The blue cross shows the position of the nucleus. The fluxes are in units of $10^{-15}$ erg cm$^{-2}$ s$^{-1}$.}
  \label{fig:maps}
\end{figure*}

\begin{figure*}
  \begin{tabular}{cc}
    \includegraphics[scale=0.47]{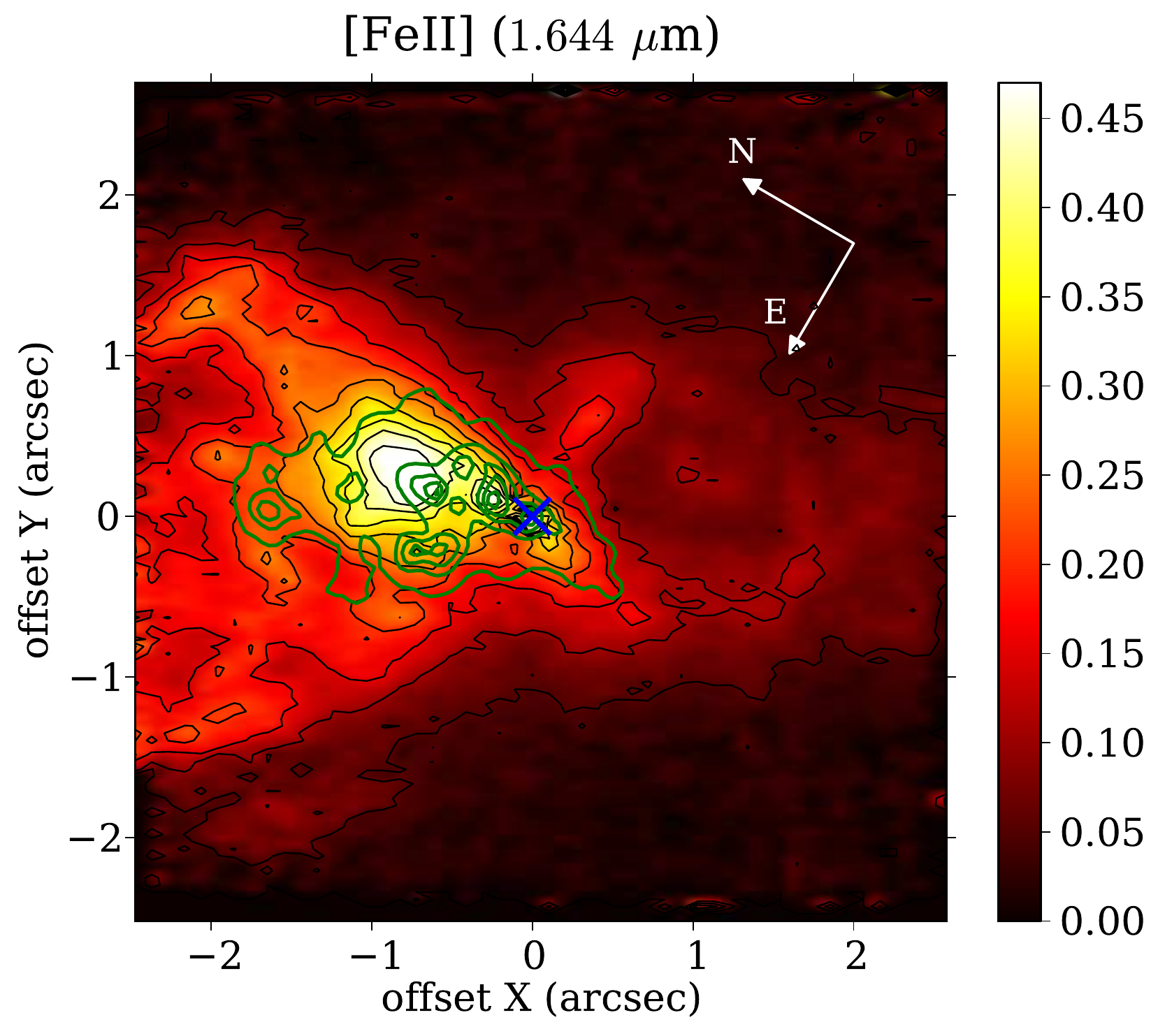}&
    \includegraphics[scale=0.47]{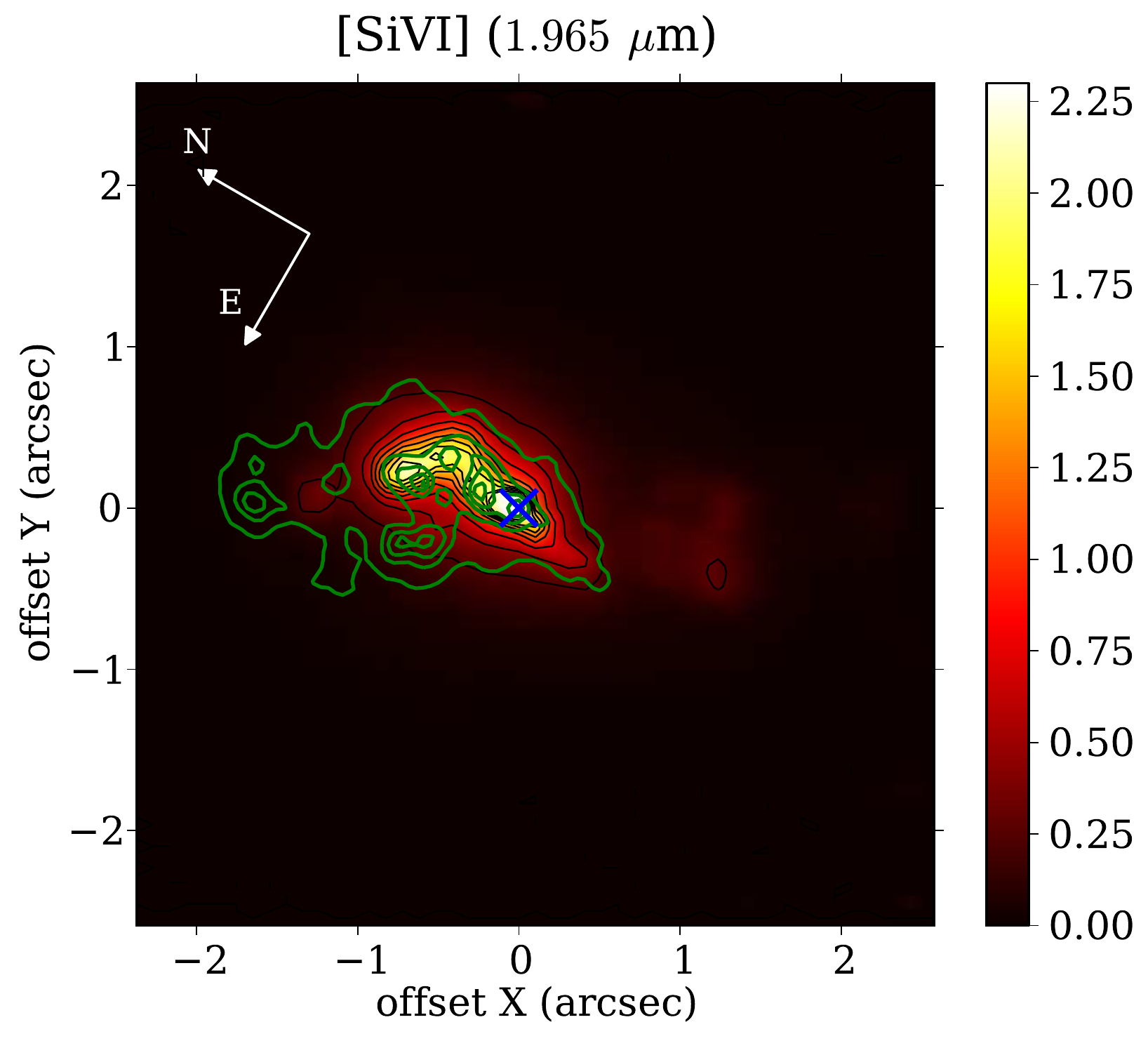} \\
    \includegraphics[scale=0.47]{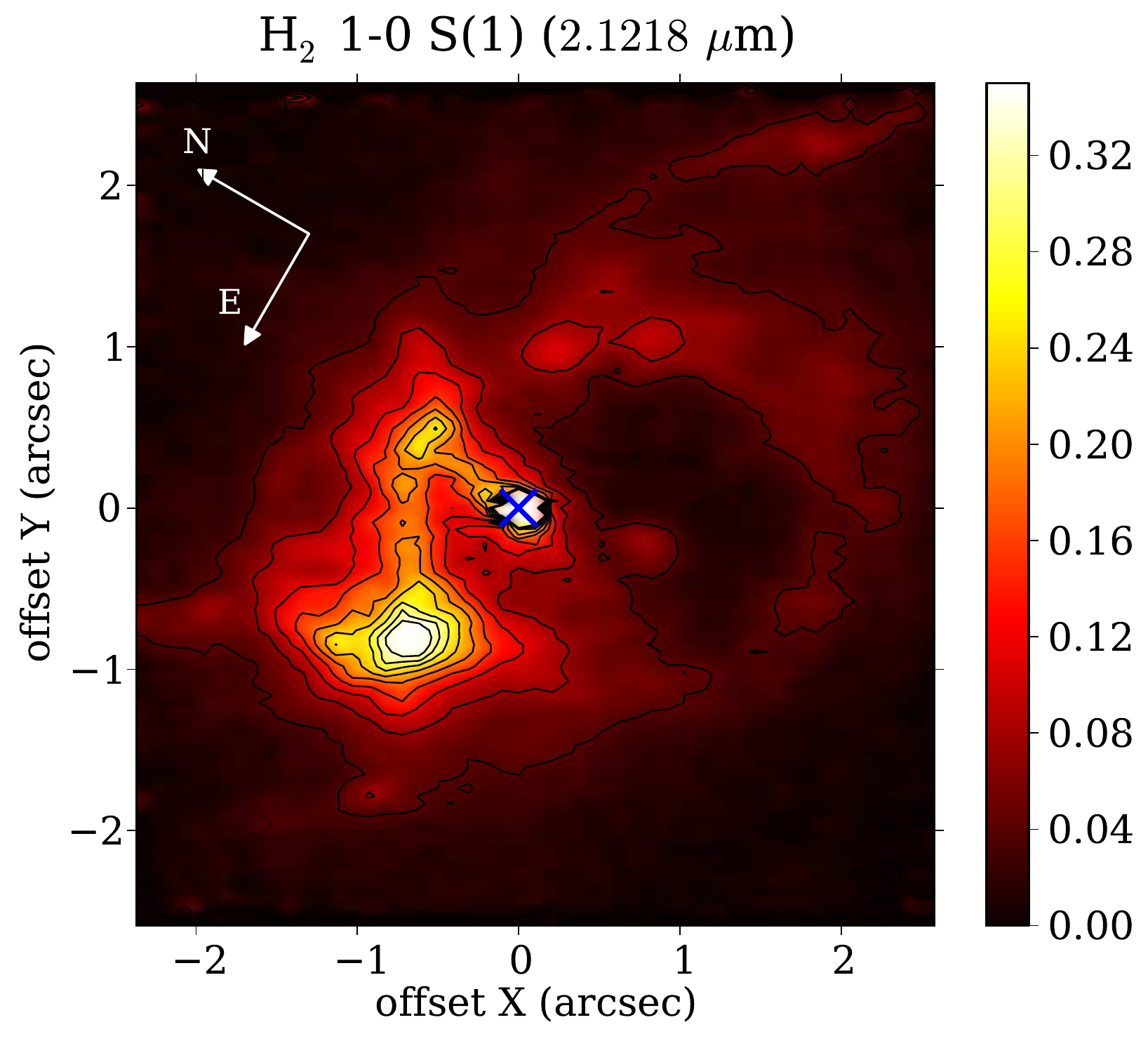} &
    \includegraphics[scale=0.47]{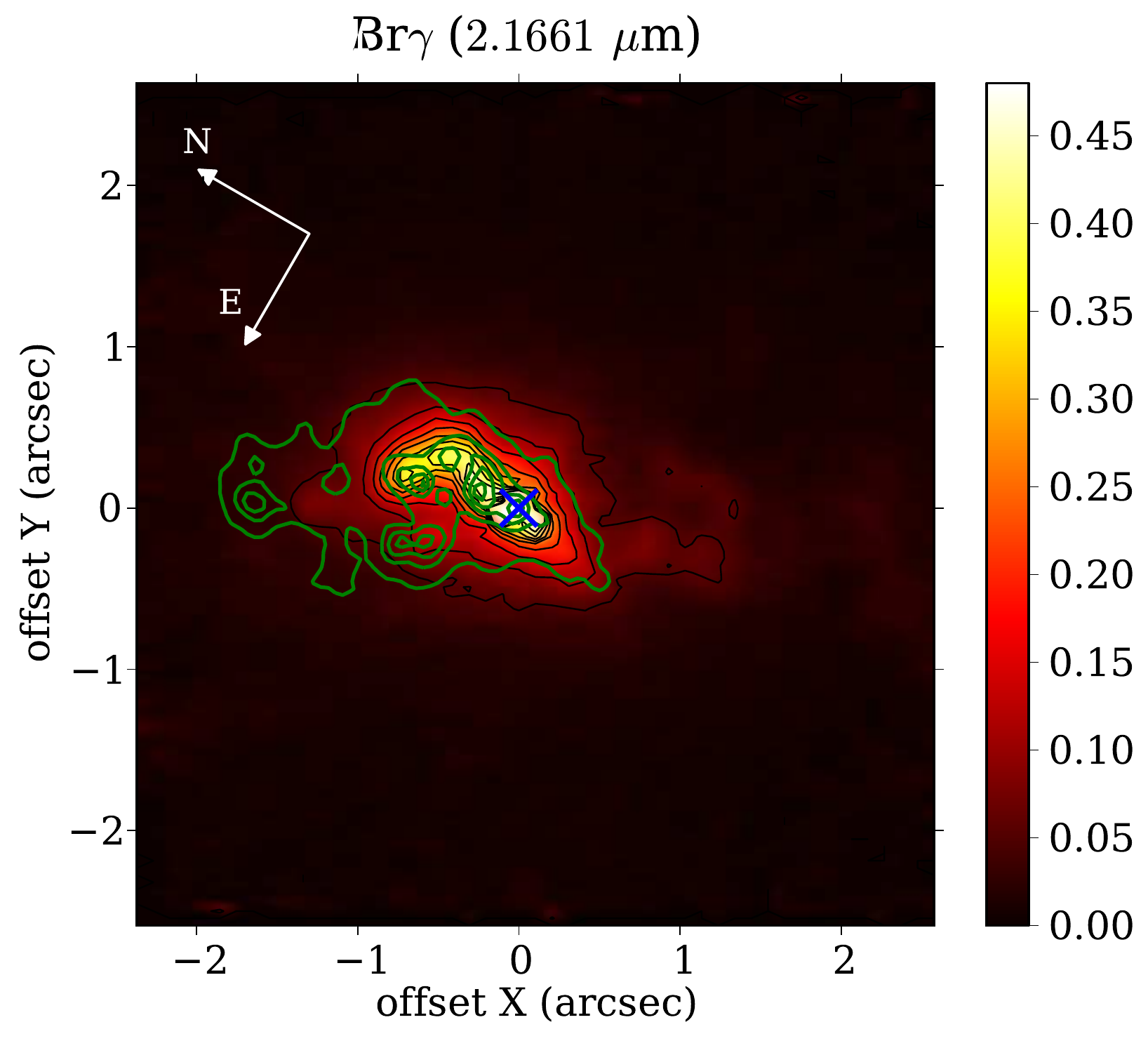} \\
    \includegraphics[scale=0.47]{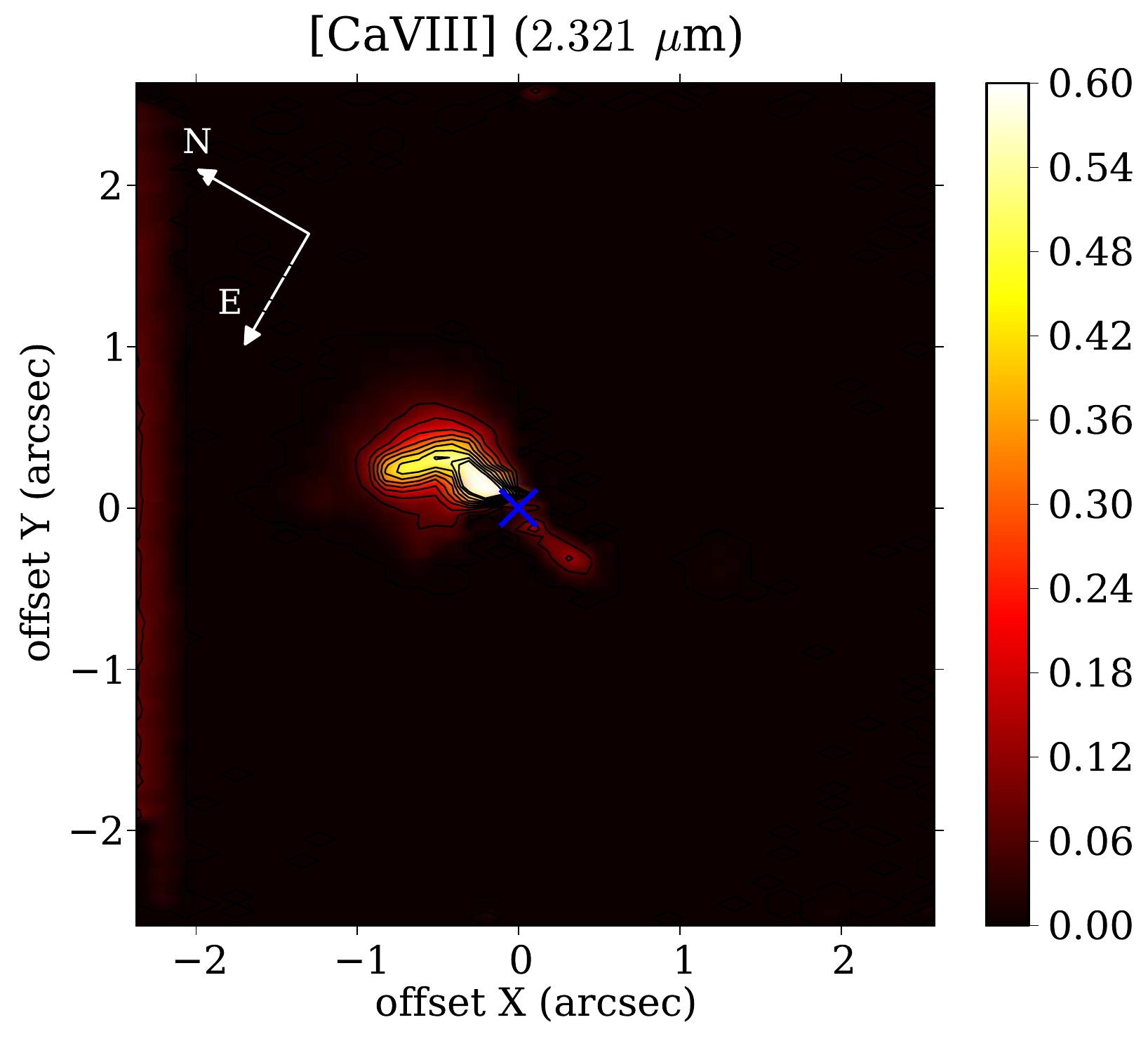}&
    \includegraphics[scale=0.47]{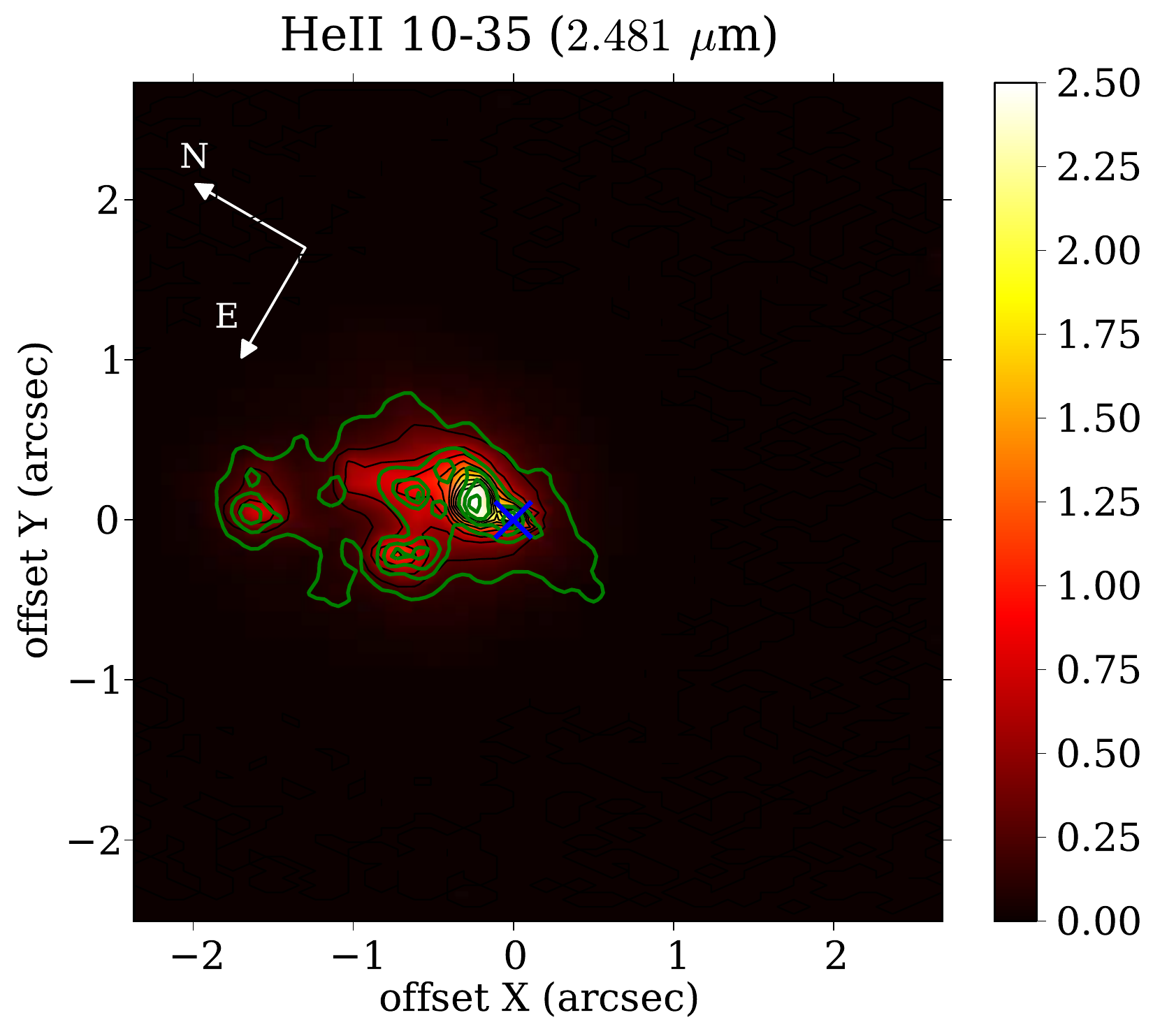}
  \end{tabular}
  \caption{Flux distributions in the H- and  K-band emission lines: [Fe{\sc\,ii}] ($1.644\,\mu$m), 
[Si{\sc\,vi}] ($1.965\,\mu$m), H$_2$ ($2.1218\,\mu$m), Br$\gamma$ ($2.1661\,\mu$m), Ca{\sc\,viii}] ($\lambda2.321\,\mu$m)  and [He{\sc\,ii}] ($2.481\,\mu$m). Contours of the 
[O{\sc\,iii}]$\lambda$5007 image from Schmitt \citep{sda2003} are overplotted in green on some of the flux maps.
The blue cross shows the position of the nucleus. The fluxes are in units of $10^{-15}$ erg cm$^{-2}$ s$^{-1}$.}
  \label{fig:mapsK}
\end{figure*}

We obtained emission-line flux distributions by integrating the flux of each emission line over the whole FOV, after subtraction of the underlying continuum contribution. In Figure\,\ref{fig:maps} and \ref{fig:mapsK} we present these maps for the emission lines with the highest signal-to-noise (S/N) ratio among their species: [P{\sc\,ii}] ($\lambda 1.1886\,\mu$m), [S{\sc\,ix}] ($\lambda 1.2523\,\mu$m), [Fe{\sc\,ii}] ($\lambda 1.257\,\mu$m), Pa$\beta$ ($\lambda 1.2821\,\mu$m), [Fe{\sc\,ii}] ($1.644\,\mu$m), [Si{\sc\,vi}] ($1.965\,\mu$m), H$_2$ ($2.1218\,\mu$m), Br$\gamma$ ($2.1661\,\mu$m), [Ca{\sc\,viii}] ($\lambda2.321\,\mu$m) and [He{\sc\,ii}] ($2.481\,\mu$m). The green contours overploted on the [Fe\,{\sc ii}] ($1.644\,\mu$m), [Si{\sc\,vi}] ($1.965\,\mu$m), Br$\gamma$ ($2.1661\,\mu$m) and [He{\sc\,ii}] ($2.481\,\mu$m) flux distributions are from the HST [O{\sc\,iii}]$\lambda$5007\AA\  narrow-band image from \citet{sda2003}. In these maps, we masked regions with fluxes smaller than the standard deviation obtained from the average ``fluxes"  of regions devoid of emission.

The [P{\sc\,ii}] ($1.189 \mu$m) flux distribution is more extended to the NE, where it reaches up to 2\farcs0 from the nucleus, while to the opposite side (SW) there is no emission farther than 1\farcs0 from the nucleus. The brightest region is located at $0\farcs5$\,N from the nucleus.  The [S{\sc\,ix}] ($1.252 \mu$m) flux distribution presents a very similar morphology to that of  [P{\sc\,ii}], but the brightest region is closer to the nucleus, at $0\farcs2$\,N.

Although the [Fe{\sc\,ii}] ($1.257 \mu$m) and [Fe{\sc\,ii}] ($1.644\mu$m) flux distributions are similar to each other (as they should be), there is ``less definition'' in the former when compared to the latter due to the fact that the image quality is inferior in the J band when compared to that in the H band (PSF width of 0\farcs14 vs. 0\farcs11): much less details are seen in the [Fe{\sc\,ii}] ($1.257 \mu$m) than in the  [Fe{\sc\,ii}] ($1.644\mu$m) flux distribution.  On the basis of the latter, the [Fe{\sc\,ii}]  flux distribution can be described as  having a bi-polar ``bowl" or  hourglass morphology, oriented along NE--SW. This morphology  is ``broader" at the apex (nucleus) than the shape of the NLR in the [O\,{\sc iii}] image, as can be seen in the left panel of Fig.\,\ref{fig:mapsK}, where the contours of the [O\,{\sc iii}] image from \citet{sda2003} are overlaid on  the [Fe{\sc\,ii}] ($1.644\mu$m) flux distribution. 
The [Fe\,{\sc ii}] emission extends up to the borders of the NIFS FOV ($\approx200$\,pc from the nucleus) along the direction of the axis of the hourglass, with higher fluxes to the NE than to the SW. At the highest flux levels, the [Fe{\sc\,ii}] flux distribution approximately follows that of  [O{\sc\,iii}]$\lambda$5007\AA, but the [Fe{\sc\,ii}] emission extends beyond the [O{\sc\,iii}] emission distribution.

The Pa$\beta$ ($1.282 \mu$m) flux distribution presents an elongated emission distribution to the N--NE, reaching its maximum approximately at $0\farcs4$ N from the nucleus. It also presents some fainter emission  to the SW. The  Br$\gamma$ ($\lambda2.1661\,\mu$m) flux distribution, although showing  a similar flux distribution to that of Pa$\beta$ (as it should have), shows a much higher definition, as the PSF is narrower in the K band than in the J band, as explained above: much more detail is seen in the Br$\gamma$ flux distribution than in the Pa$\beta$ one. The Br$\gamma$ map is much more similar to the  [O\,{\sc iii}] map than the [Fe{\sc\,ii}] map: individual ``emission knots'' seen in the [O\,{\sc iii}] flux distribution are also observed in the Br$\gamma$ map. The Br$\gamma$ flux distribution is also similar to those of the [Si{\sc\,vi}] ($\lambda1.965\,\mu$m), [Ca{\sc\,viii}] ($\lambda2.321\,\mu$m) and [He{\sc\,ii}] ($\lambda2.481\,\mu$m) flux distributions.  Common to these two flux distributions are the orientation and morphology, which are also co-spatial with that of [O{\sc\,iii}], extending mostly to the N, in the region where the [Fe{\sc\,ii}] emission is also brightest. Compared to [Si{\sc\,iv}], [Ca{\sc\,viii}] and Br$\gamma$, the He{\sc\,ii} emission presents two additional knots: one at $\approx\,1\farcs6$\,NE of the nucleus (PA=30$\degr$) and the other at $\approx\,0\farcs6$\,E (PA=80$\degr$). These knots are also present  in the [O\,{\sc iii}] flux distributions and are observed in all other He{\sc\,i} and He{\sc\,ii} flux distributions.

The flux distribution of the molecular gas H$_2$ ($\lambda2.1218\,\mu$m) is completely distinct from that of the ionized gas, being distributed in a circumnuclear ring with a diameter of $\approx\,150$pc (2\farcs0). Although this ring is circumnuclear, it is not centered at the nucleus. The center of the ring is displaced towards a small knot of H$_2$ emission observed at $0\farcs8$ SW of the nucleus. All the remainder H$_2$ emission lines present in the K-band spectra show similar flux distributions.

The coronal lines flux distributions were already shown and discussed by \citet{mazzalay13b} using the same NIFS data presented here. We only show again the [S{\sc\,ix}] ($1.2523\mu$m) and [Si{\sc\,vi}] ($1.965\,\mu$m) flux distributions for the purpose of comparing them with those for the low-ionization and molecular gas. Their flux distributions are very similar to those of Pa$\beta$ and Br$\gamma$.

\subsection{Emission-line ratio maps}
\label{subsec:line_ratios}

\begin{figure*}
    \begin{tabular}{cc}
	\includegraphics[scale=0.45]{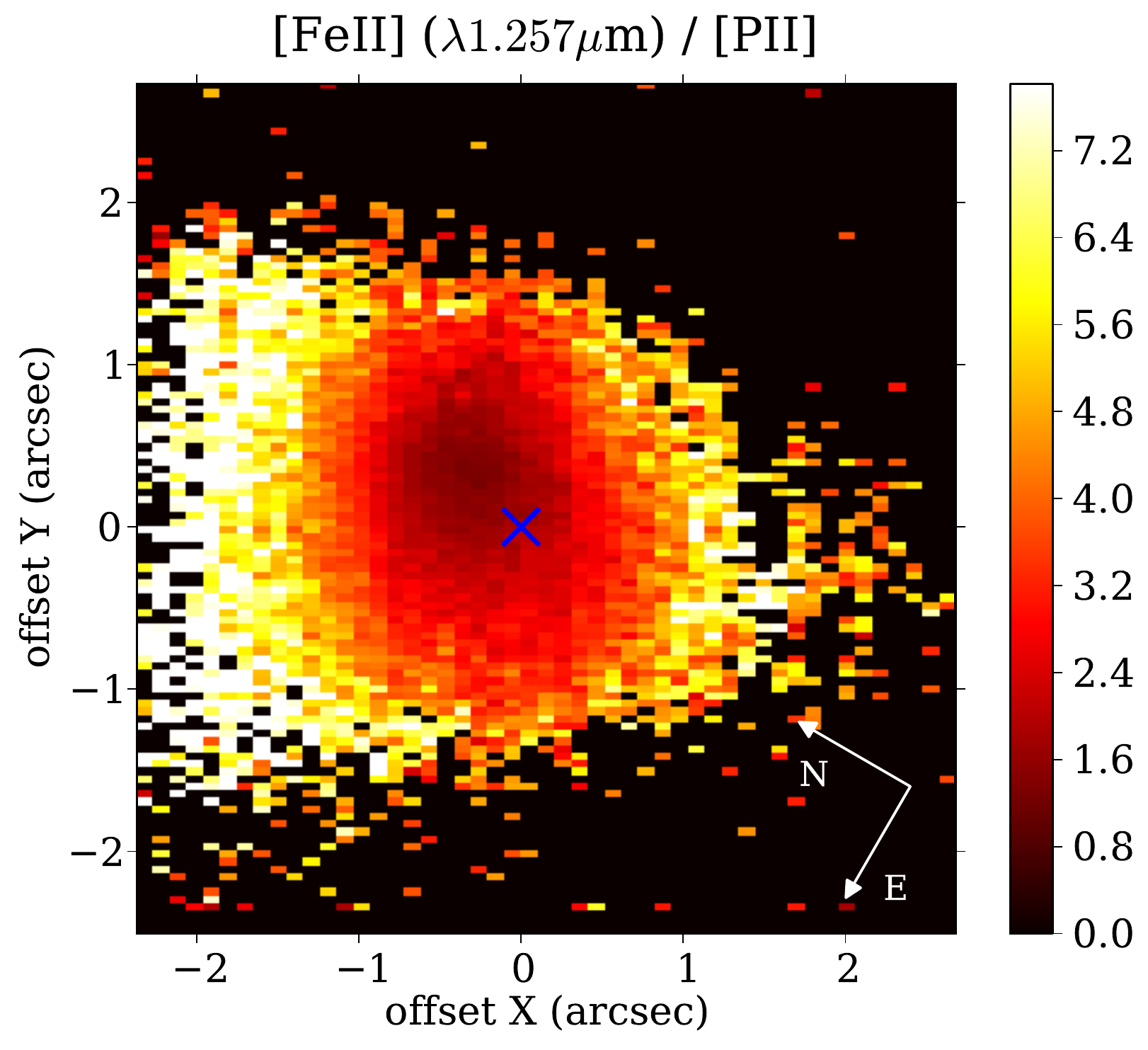} &
	\includegraphics[scale=0.45]{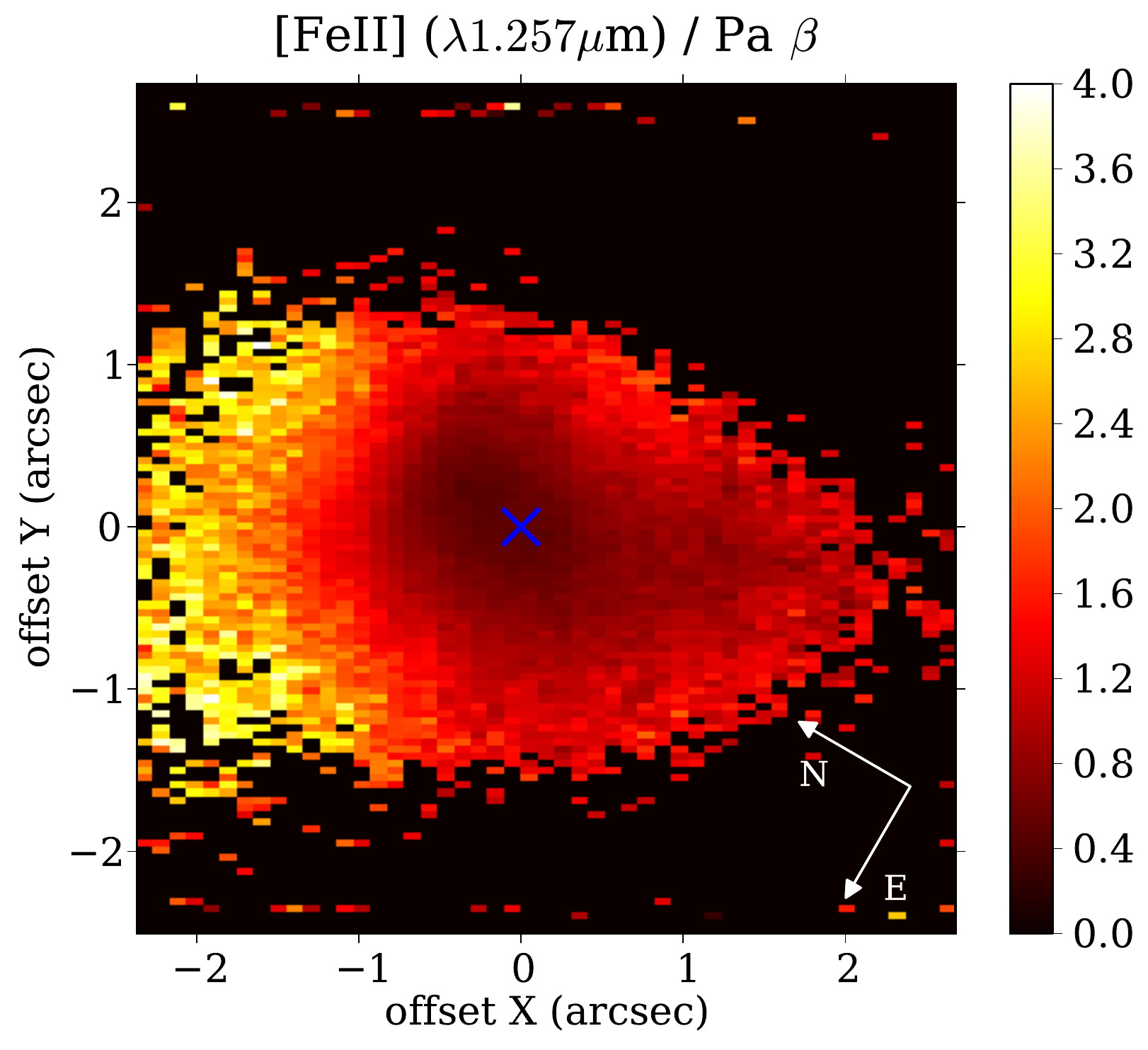} \\
	\includegraphics[scale=0.45]{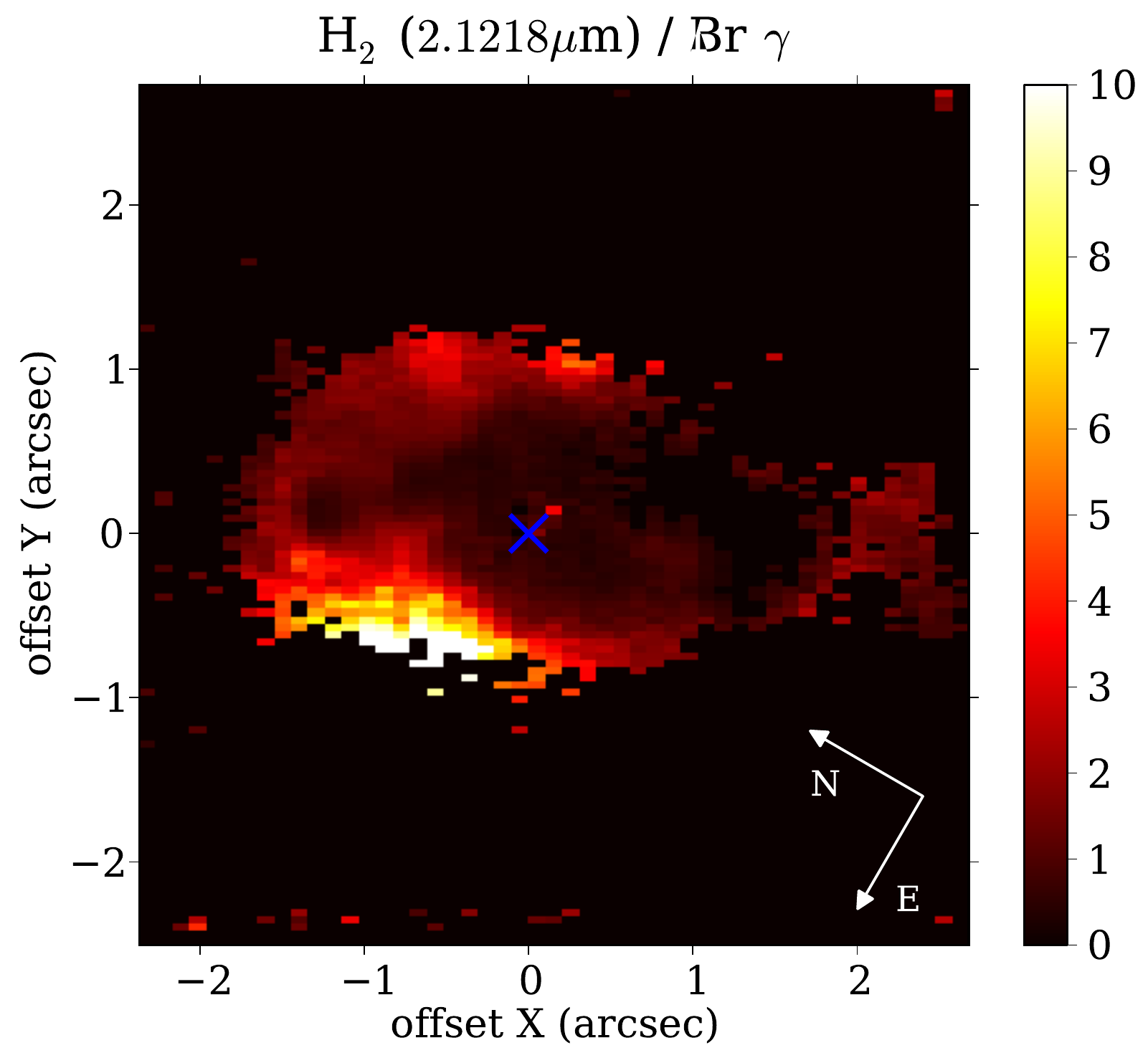}&
	\includegraphics[scale=0.45]{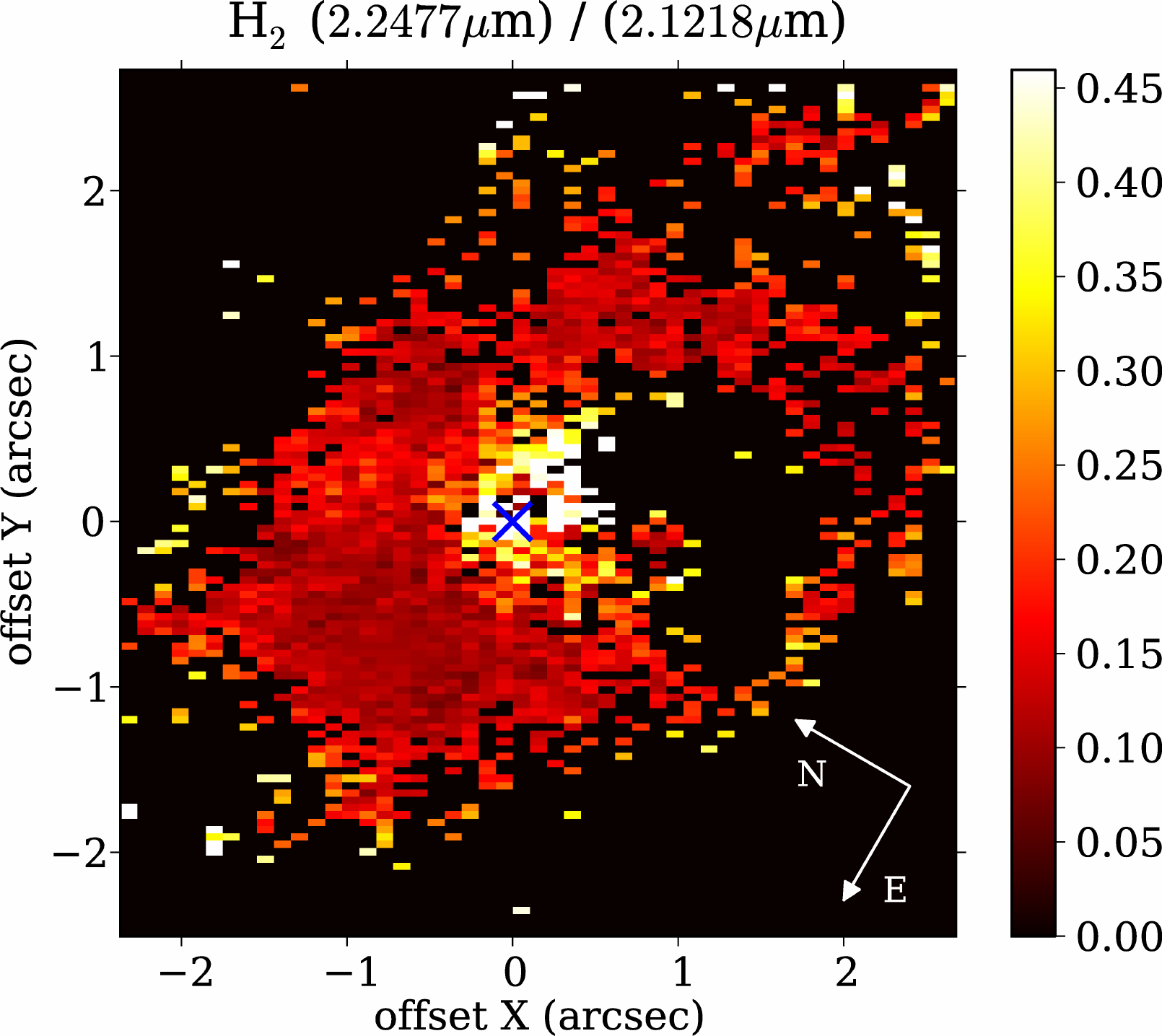} \\
    \end{tabular}
    \caption{Line-ratio maps. Top-left: [Fe{\sc\,ii}]$1.257$/[P{\sc\,ii}]$1.118$; top-right: [Fe{\sc\,ii}]$\lambda1.257$/Pa$\beta$; bottom-left: H$_2 2.1218/{\rm Br}\gamma$ and bottom-right: H$_2$\,2.2477/H$_2$\,2.1218. The blue cross indicates the position of the nucleus.}
    \label{fig:ratios}
\end{figure*}

We have used the flux distributions to create the following line-ratio maps: [Fe{\sc\,ii}]$\lambda1.257$/[P{\sc\,ii}]$\lambda1.1886$, [Fe{\sc\,ii}]$\lambda1.257$/Pa$\beta$, H$_2 2.248/2.122$ and H$_2 2.1218/{\rm Br}\gamma$. These maps are shown in Fig. \ref{fig:ratios} together with reddening $E(B-V)$ maps derived  from the Br$\gamma$/Pa$\beta$ and [Fe{\sc\,ii}]$\lambda1.257/1.644$ ratios. These ratios can be used to investigate the excitation mechanisms of the $H_2$ and [Fe\,{\sc ii}] near-IR emission lines and to map the extinction in the NLR. As for the case of the flux distribution maps, we have masked regions where the S/N ratio was not high enough to fit the emission-line profiles and excluded these regions from the ratio maps.

In order to compute ratios between lines from H or K  band and the those from the J band, we have convolved the best resolution images -- from the H and K bands -- with a Gaussian kernel with $\sigma$ equivalent to $0\farcs3\times0\farcs3$ [5 pixels along the slitlets (y) $\times$ 2 pixels across slitlets (x)]. In order to avoid infinities, we have masked out from the ratio maps any pixel for which the flux was smaller than the standard deviation of the flux in a region devoid of emission ("sky").

In the top-left panel of Fig.~\ref{fig:ratios} we show the [Fe{\sc\,ii}]$1.257$/[P{\sc\,ii}]$1.118$ line-ratio map, which is useful to investigate the origin of the [Fe\,{\sc ii}] line emission; higher values correspond to a larger contribution from shocks to the excitation of the [Fe\,{\sc ii}] \citep[e.g.][]{oliva01,sb09,mrk1066a}. The smallest values of [Fe{\sc\,ii}]$1.257$/[P{\sc\,ii}]$1.118\approx$1.3 are observed at $0\farcs4$ N from the nucleus, while the highest values of up to 10 are seen at the borders of the IFU field to the northeast. Some high values of this ratio are also seen in a small region  at 1\farcs5 southwest of the nucleus.  

The [Fe{\sc\,ii}]$\lambda1.257$/Pa$\beta$ line ratio map is shown in the top-right panel of Fig.~\ref{fig:ratios}. It shows a similar structure to that observed in the [Fe{\sc\,ii}]$1.257$/[P{\sc\,ii}]$1.118$ ratio map, with the smallest ($<$0.6) values being observed at $0\farcs4$ N from the nucleus and the highest values of up to 3  to the northeast, near  the border of the FOV.

The bottom-left panel of Fig.~\ref{fig:ratios} presents the H$_2 2.1218/{\rm Br}\gamma$ line-ratio map. At most locations this ratio present values ranging from 0.6 to 2, with the smallest values observed  at  at $0\farcs4$ N from the nucleus -- where the flux distributions from the ionized gas have a peak. Nevertheless,  values of up to 10 are observed 
at 1$^{\prime\prime}$\,NE from the nucleus, in a region where the H$_2$ emission presents a knot of enhanced emission and Br$\gamma$ is very weak.

Finally, the bottom-right panel of Fig.~\ref{fig:ratios} shows the H$_2$\,2.2477/H$_2$\,2.1218 line-ratio map. Its highest value of 4.7 is observed at the nucleus, as well as at its surroundings where this line-ratio reaches $\approx$\,2.0 up to $\approx 0\farcs4$ W of the nucleus. Along the H$_2$ ring, the values are $\approx$\,0.11, while in regions at the borders of the ring this line ratio increases to about 0.35.

\section{Discussion}
\label{sec:discussion}

\subsection{Emission-Line Profiles}

 The emission-line profiles over most of the observed field-of-view are complex, presenting double or multiple components originated in gas with distinct kinematics. These complex profiles can be seen in Fig.~\ref{fig:sample_spectra}: the Pa$\beta$ profile is double peaked in positions B and C as well as the [Fe\,{\sc ii}] profile in position B. The study of each kinematic component individually would likely yield further insights into the physical conditions, reddening, and geometry of the NLR, but this is a task beyond the scope of this paper. A detailed analysis of the gas kinematics is presented in a companion paper \citep{barbosa14}. In the present paper, we do not separate the different kinematic components and use instead the total flux in the lines to map the average extinction and gas excitation.

At the nucleus and surrounding regions, the Pa$\beta$ and Br$\gamma$ emission-line profiles present a broad ``base" that could be interpreted as coming from the broad line region (BLR). In the case of Pa$\beta$, we could fit a narrow and a broad component with FWHM$\approx$1500~\kms\ and a full width at zero intensity -- FWZI$\approx$4700\,\kms; could this come from the BLR? The BLR in NGC~1068 was observed for the first time in polarized light  \citep[e.g.][]{antonucci1985,simpson02}, with the emission lines showing  FWZI$\approx$7500\,\kms and attributed to emission from the BLR scattered towards the observer. This width is approximately twice the value we measure for the broad base of the nuclear Pa$\beta$ emission line. This fact, combined with the observation that as one moves away from the nucleus the profile approximately keeps its width but shows more clearly separate kinematic components, suggests that the broad component we observe in Pa$\beta$ is actually a superposition of two or more kinematic components of the NLR gas, instead of from gas of the BLR. This is also supported by the presence of multiple components in the Pa$\beta$ profile at the position B (at the location of the peak of the [Fe\,{\sc ii}] emission). At this location the Pa$\beta$ profile is clearly double peaked with a similar FWZI to that of the nuclear profile. This seems to apply also to the Br$\gamma$ profile but is less clear due to the lower intensity of the line.


\subsection{Extinction}
\label{subsec:extinction}

\begin{figure*}
    \centering
    \begin{tabular}{cc}
	\includegraphics[scale=0.45]{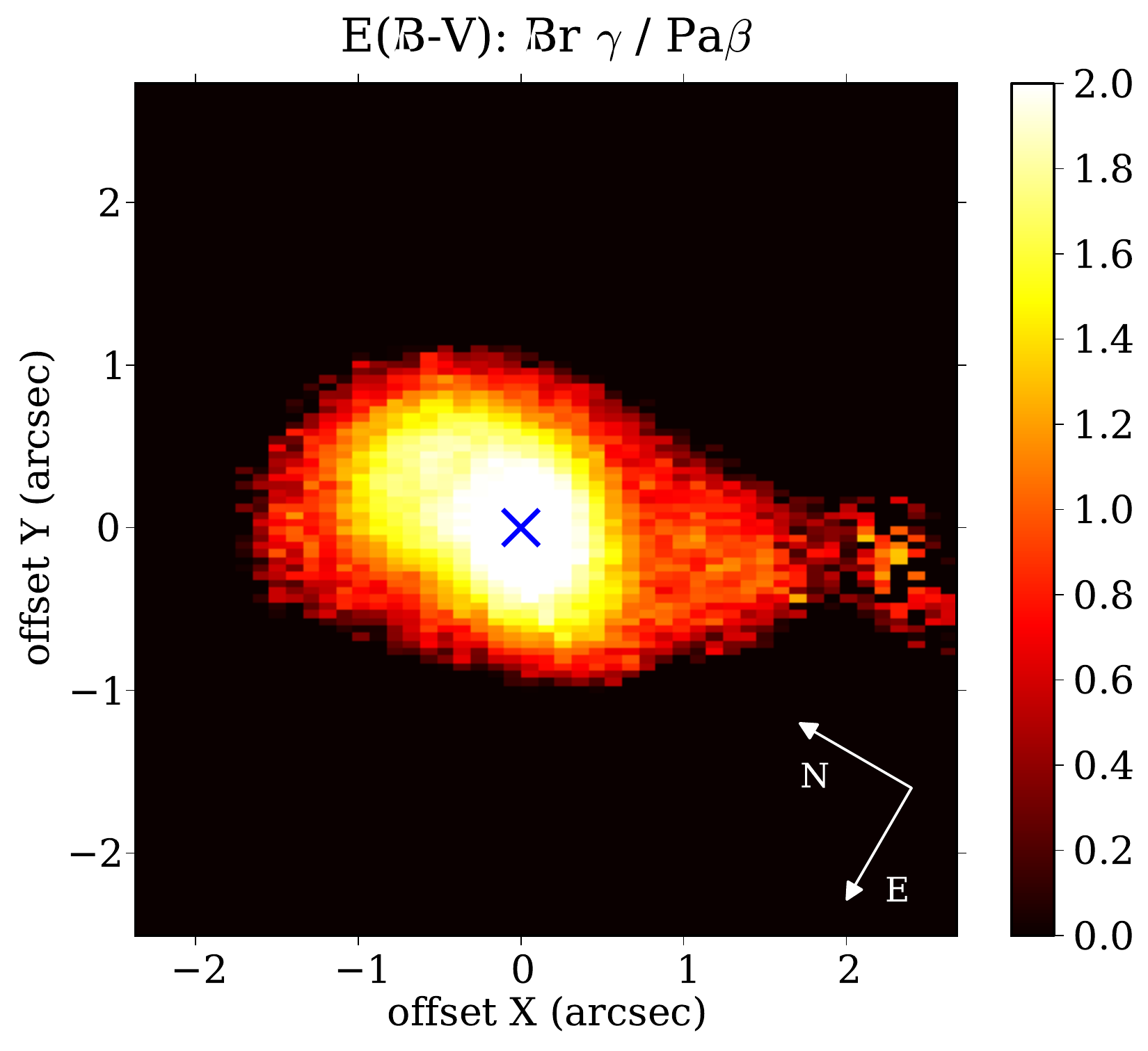}&
	\includegraphics[scale=0.45]{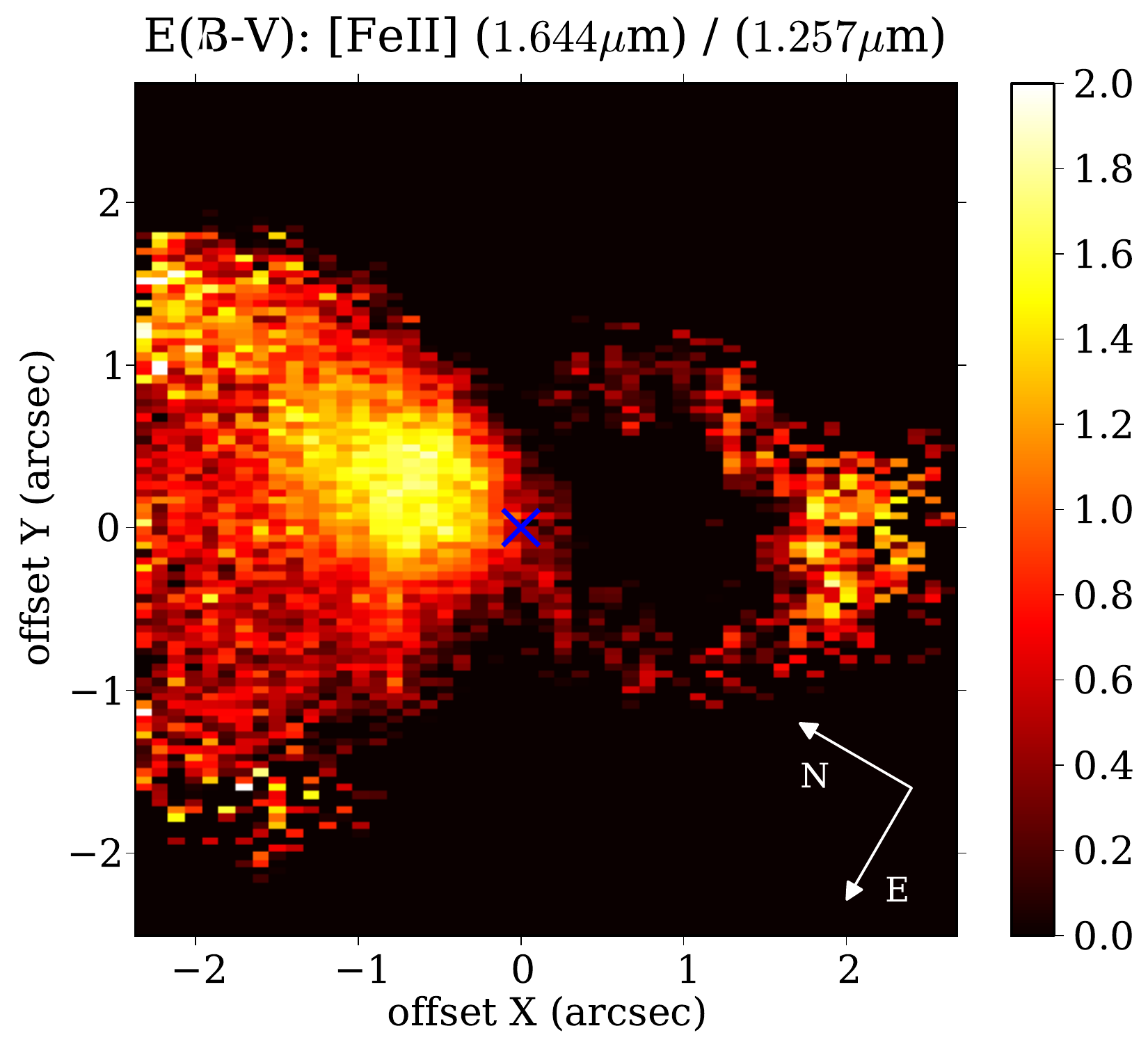}
    \end{tabular}
    \caption{Average extinction E(B-V) obtained from the ratios Br$\gamma$/Pa$\beta$ (left panel) and [Fe{\sc\,ii}]$\lambda$1.644/1.257 (right panel). The blue cross shows the position of the nucleus.}
    \label{fig:ebv}
\end{figure*}

In order to map the NLR extinction, we used the  Br$\gamma$/Pa$\beta$ and [Fe{\sc\,ii}]$\lambda1.257/1.644$ emission-line ratios to derive the average  reddening of  the emitting gas. Following \citet{sb09}, the reddening can be obtained from the Br$\gamma$/Pa$\beta$ ratio using:

\begin{equation}
 E(B-V)=4.74\,{\rm log}\left(\frac{5.88}{F_{Pa\beta}/F_{Br\gamma}}\right),
\end{equation}
where $F_{Pa\beta}$ and $F_{Br\gamma}$ are the fluxes of $Pa\beta$ and $Br\gamma$ emission lines, respectively. We have used the reddening law of \citet{cardelli89} and have adopted the intrinsic ratio $F_{Pa\beta}/F_{Br\gamma}=5.88$ corresponding to case B recombination \citep{osterbrock06}. The resulting $E(B-V)$ map is shown in the left panel of Fig.\,\ref{fig:ebv}. This map shows that the highest $E(B-V)$ values reach up to 1.8 mag, and outline a structure elongated to the north of the nucleus and extending to $\approx$1\farcs0 from it. In regions away from this structure, a typical average value is $E(B-V)\approx$0.8 mag.

The [Fe{\sc\,ii}]$\lambda1.257$ and [Fe\,{\sc ii}$\lambda1.644$ emission lines arise from the same upper level with an intrinsic ratio of [Fe{\sc\,ii}]$\lambda1.257/1.644 =1.36$ \citep[e.g.][]{nussbaumer88,bautista98,sb09}. Following \citet{sb09}, an $E(B-V)$ map can also be obtained from:

\begin{equation}
E(B-V)=8.14\times log\left(\frac{1.36}{F_{1.2570}/F_{1.6440}}\right),
\end{equation}
using again the reddening law of \citet{cardelli89}. The resulting reddening map is shown in the right panel Fig.~\ref{fig:ebv}, which shows values consistent with those obtained from the Br$\gamma$/Pa$\beta$ line ratio, even though the [Fe\,II] reddening map is more extended due to the fact that the [Fe\,II] maps are more extended than those of Pa$\beta$ and Br$\gamma$. The exception is the nucleus, where $E(B-V)_{[Fe\,II]}\approx$\,0.6\,mag is smaller than $E(B-V)_{Br\gamma/Pa\beta}\,\approx$\,1.8\,mag. This difference could be due to contamination of the H lines, and specially of Br$\gamma$ -- which is not easily measurable at the nucleus -- by a broad component. Alternatively, as the [Fe\,II] lines originate in partially ionized regions, that are farther from the source than the fully ionized regions, the average extinction probed by these lines could be smaller than that probed by the H lines.

Common to both E(B-V) maps are: (1) the region of highest reddening, which is observed from the nucleus up to 1\farcs5 close to the N wall of the cone (or hourglass), and (2)  the region at $\approx$2\farcs0 southwest of the nucleus, where the reddening values are similar in both maps. The other regions of the NLR have average $E(B-V)$ values of 0.6.

We can compare the $E(B-V)$ obtained here for the NLR of NGC\,1068 with those from previous works.
\citet{koski78} reported a value of $E(B-V) = 0.52$~mag from integrated spectra within an aperture of $2\farcs7\times4\arcsec$, while \citet{ho97} obtained $E(B-V) = 0.54$~mag as estimated from the Balmer decrement from observations within a similar aperture; this aperture corresponds to integrating our data over approximately half our field-of-view. These values are consistent with our average value of E(B-V)$\approx$0.6, observed over most of the region from where we could measure it from the [Fe\,II] line ratio. \citet{kraemer2000} used HST STIS spectra with a slit width of 0\farcs1 with the slit centered on a position 0\farcs14 north of the optical continuum peak oriented along PA=202$^\circ$ and calculated the reddening along the slit using the He\,{\sc ii}$\lambda\lambda1640,4686$  emission lines. They show that the reddening along the slit is highly variable, obtaining values in the range $0.1\,\le$\,E(B-V)$\le$\,0.6 and estimate an average reddening of E(B-V)$\approx$0.35 for the blueshifted emitting gas to the NE and of E(B-V)$\approx$0.22 for the redshifted emitting gas to NE. These values are smaller than those we have obtained. Our data in the J band ([Fe{\sc\,ii}]$\lambda1.257$) does not have the angular resolution of these HST observations, and we thus cannot sample variations on this fine scale, but our average E(B-V) values are higher than those obtained from the He\,{\sc ii} lines. One possible explanation is that the lower-ionization species (as H~{\sc ii} with ionization potential $IP=13.6~$eV and Fe~{\sc ii} with $IP=7.87~$eV) we are observing originate in regions of higher extinction than those from where the the high-ionization gas emission (such as He~{\sc ii}, with $IP=24.6$~eV)  originates.
Using near-IR long-slit spectra, although with ground-based $\sim$1\arcsec angular resolution \citet{martins10} reported values of $E(B-V)$ ranging from 0.1 to 2.65 along a 0\farcs8 width slit covering the inner 5$^{\prime \prime}$ oriented along the N-S direction, also derived from Pa$\beta/$Pa$\gamma$ and [Fe\,{\sc ii}]1.257/1.643 line ratios. For the inner 2\farcs5, \citet{martins10} also obtained larger values for $E(B-V)$ derived from H recombination lines than from the [Fe\,{\sc ii}] lines, similarly to what we have found. For the nucleus, these authors found $E(B-V)=1.13\pm0.10$ and $E(B-V)=0.26\pm0.12$ based on H and [Fe~{\sc ii}] lines, respectivelly.  Our value of  $E(B-V)$ estimated from the H lines for the nucleus is slightly larger than that of \citet{martins10}, possibly due to the uncertainties in the fitting of the Br$\gamma$ emission line discussed above. Considering that our angular resolution is higher than that of \citet{martins10}, their value can be considered an average of our values integrated within 0\farcs8$\times$1\farcs6 (their nuclear aperture). In summary, the reddening for the nucleus derived using near-IR H lines is higher than both those derived from H optical lines as well as from those derived from the [Fe{\sc\,ii}] near-IR lines. Either the near-IR H lines are probing regions of higher reddening or there could be some contribution from a broad Br$\gamma$ component  that increases the nuclear Br$\gamma$/Pa$\beta$ ratio. Outside the nucleus, the reddening values are more similar, and the average value $E(B-V)\,\approx\,0.6$ is not that different from that derived from optical emission lines. 



%
%
\subsection{Ionized and molecular gas distributions}
\label{subsec:dist}
\subsubsection{Ionized gas distributions}
\label{subsec:dist-ionized}

\begin{figure}
  \begin{tabular}{cc}
    \includegraphics[scale=0.47]{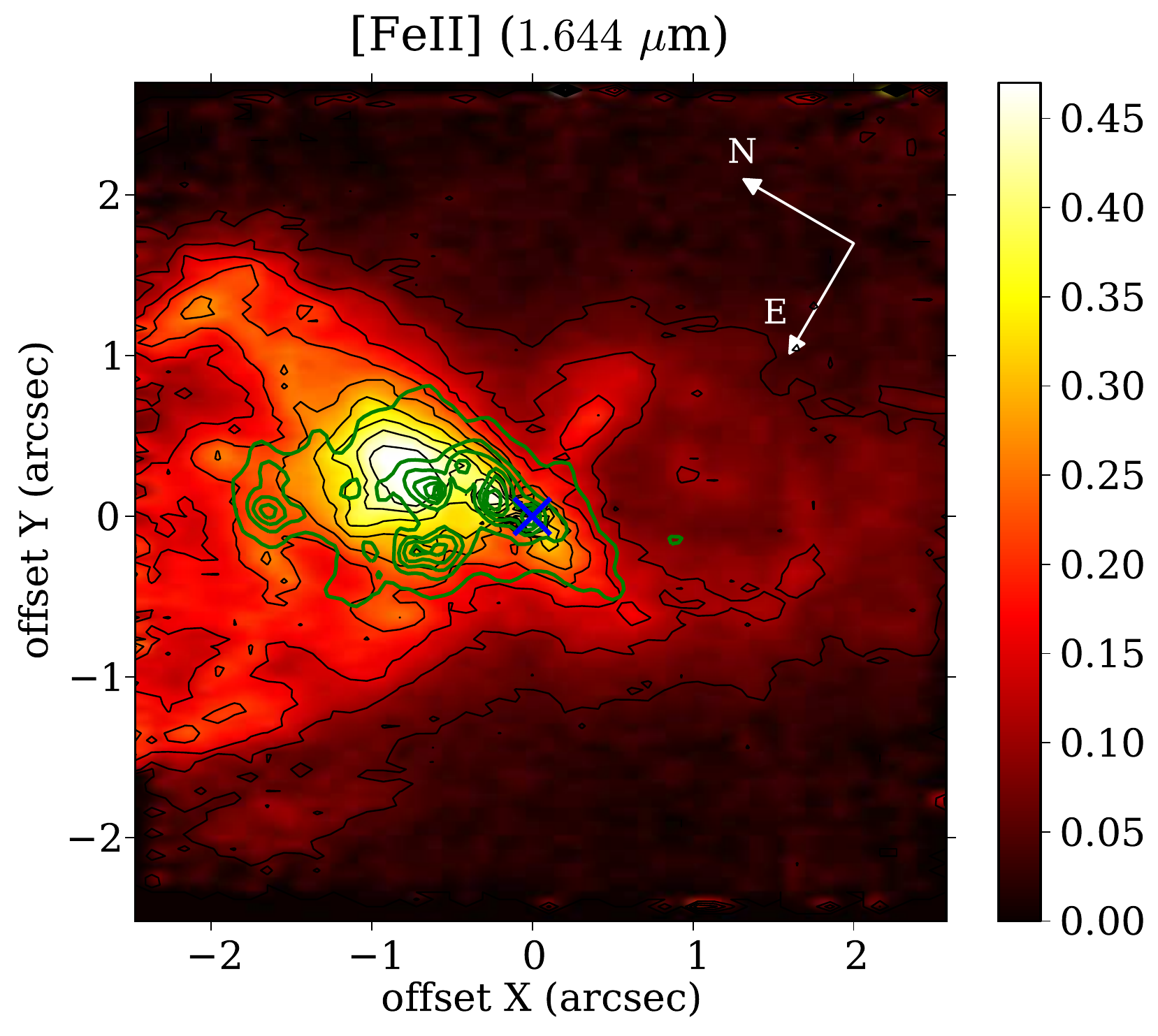}&
  \end{tabular}
  \caption{Comparison between the [Fe{\sc\,ii}] ($1.644\,\mu$m) and [O\,{\sc iii}] flux distributions. [O\,{\sc iii}] image contours from \citet{sda2003} are shown in green with levels ranging from 2$\sigma$ to the maximum of the [O\,{\sc iii}] emission flux ($\sim 30\sigma$) with an interval of 5$\sigma$ between  contours. 
The blue cross shows the position of the nucleus. The fluxes are in units of $10^{-15}$ erg cm$^{-2}$ s$^{-1}$.}
  \label{fig:compFe}
\end{figure}

In Sec.\,\ref{subsec:emission} we have compared the emission-line flux distributions shown in Fig.~\ref{fig:maps} with the optical [O\,{\sc iii}]$\lambda$5007\AA\ flux distribution. Most ionized gas flux distributions  resemble that of the [O{\sc\,iii}]$\lambda 5007$\AA\,emission line. All knots seen in the [O\,{\sc iii}] contour map are seen also in the Br$\gamma$ and  He\,{\sc ii} flux distributions, for example.  The NE side of the NLR is brighter than the SW side in all flux distributions of the ionized gas, as also observed in [O\,{\sc iii}], what has been attributed to the orientation of the bipolar outflow, which is in front of the galactic plane to the NE and behind the plane to the SW \citep{das06}. 
This resemblance of the near-IR ionized gas and [O\,{\sc iii}] flux distributions suggests a similar origin for the corresponding emission lines, which previous works have attributed to emission from the walls of a hollow bi-cone centered in the nucleus and oriented along PA$=30/210$\deg, as described by \citet{das06,das07}. But although most of the ionized gas flux distributions show several resemblances with that of the [O\,{\sc iii}], some differences can be observed in the case of [Fe\,{\sc ii}]. 

In order to better compare the [Fe\,{\sc ii}] and [O\,{\sc iii}] images, we show in Figure~\ref{fig:compFe} the [Fe{\sc\,ii}] ($1.644\,\mu$m) flux distribution with [O\,{\sc iii}] contours overlaid with levels ranging from 2$\sigma$ to the maximum of the [O\,{\sc iii}] emission, where the $\sigma$ is defined as the standard deviation of the fluxes in a region with no [O\,{\sc iii}] emission in the image.  From this comparison, it is clear that the [Fe\,{\sc ii}] emission is more extended than the [O\,{\sc iii}] one, with a structure better described as a section of an hourglass or of two ``bowls", one to each side of the nucleus rather than a bi-cone. This morphology is similar to that observed for some planetary nebulae (e.g. NGC\,6302). In \citet{barbosa14}, we use a ``lemniscata" geometry (similar to that of an hourglass) to model the  [Fe\,{\sc ii}] kinematics (outflow), showing that it reproduces better the flux distributions in channel maps than a bicone. We attribute the larger extent and ``broader" flux distribution to the fact that  the ion Fe\,{\sc ii} is also formed in partially ionized regions, as Fe\,{\sc i} has an ionization potential of only 7.9\,eV  \citep{worden84}. Thus, [Fe\,{\sc ii}] emission can be observed beyond the fully ionized regions, in locations where there is no more H$^+$  or O$^{++}$ but may still be outflows. For this reason, we argue that Fe$^+$ is a better tracer of outflows than H$^+$ and  O$^{++}$.


\subsubsection{Molecular gas flux distribution}

The flux distribution in the H$_2$ ($\lambda2.1218\,\mu$m) emission line shows a totally different morphology from that of the ionized gas (see Fig.~\ref{fig:maps}). The H$_2$ emission  is mostly originated in a ring with approximate major ($2a$) and minor ($2b$) axes of 3\farcs0  and 2\farcs3, respectively, measured from the bottom-left panel of Fig.~\ref{fig:maps}. The major axis of the ring is oriented approximately along the line of nodes of the galaxy, as seen from the stellar velocity field shown in \citet{sb12}.  Assuming that this ring is circular and is in the plane of the galaxy, it has a diameter of $\approx$\,200\,pc, and we can use its apparent axes diameters to estimate its inclination relative to the plane of the sky as $i={\rm cos^{-1}}(b/a)=\approx40^\circ$. This value is in approximate agreement with that of the large scale disk, derived from the extent of the major and minor axes of the galaxy quoted in  NASA/IPAC Extragalactic Database (NED)\footnote{http://ned.ipac.caltech.edu} and in \citet{devaucouleurs1991}. This inclination is also similar to that derived from our modeling of the stellar velocity field in the inner 200\,pc, which is $i\approx33^\circ$. 

The presence of the H$_2$ ring had already been inferred in previous near-IR long-slit spectroscopy with the Very Large Telescope (VLT) of the inner 1\farcs5$\times$3\farcs5 by  \citet{alloin01}. These authors concluded that  the H$_2$ emitting gas in NGC\,1068 was mainly concentrated in two knots along the East-West direction at a distance of about 70~pc from the nucleus, consistent with a location in the ring. \citet{galliano02} showed that the H$_2$ kinematics in the inner 200~pc cannot be reproduced by a simple rotating disk and suggest an additional outflowing component. They also conclude that the H$_2$ is originated in gas excited by X-ray irradiation from the central AGN.

The H$_2$ flux distribution and kinematics was also recently studied by \citet{ms09} using adaptive optics integral field spectroscopy with the VLT instrument SINFONI. They show a map for the H$_2$ (2.1218$\mu$m) flux distribution over the inner $4^{\prime \prime}\times4^{\prime \prime}$ of NGC\,1068, which is very similar to ours. They report also the detection of two prominent linear structures to the the north and south ob the nucleus observed within the inner 0\farcs4 between the ring and the AGN, which they attribute to emission from gas streaming towards the nucleus on highly elliptical or parabolic trajectories in the plane of the galaxy. 

In millimeter wavelengths, \citet{schinnerer00} present a map for the $^{12}$CO(2-1) emission, which shows that the cold molecular gas presents the same ring-like structure seen in the ``warm" H$_2$ flux distribution that we observe in the near-IR. The presence of this molecular ring is also confirmed by recent observations of emission lines from CO, HCN and HCN$^+$ \citep{krips11}. These authors show that the molecular ring is expanding outwards in the galactic plane and propose that the expansion is  caused by shocks originated in the interaction of the AGN jet with the molecular gas. In \citet{barbosa14} we show that the kinematics of the H$_2$ observed with NIFS is consistent with expansion in the plane, but most of  the AGN outflow observed in ionized gas seems not to be co-spatial with the H$_2$ ring. There is even some ``empty space" between the ionized gas emission and the H$_2$ ring observed towards the NW and SE, which do not support an interaction between the outflow and the ring to cause its expansion. The radio jet is much more collimated than the outflow and probably leaves the galaxy plane at the nucleus at an angle of at least 45\deg, and it is hard to believe that such a narrow jet would be responsible for the expansion of the ring $\approx$\,100\,pc away in the galaxy plane. In addition, in \citet{sb12}, we found that there is a correlation between the spatial distribution of a young ($\approx$\,30\,Myr) stellar population and the flux distribution of H$_2$, suggesting that the H$_2$ expanding kinematics could be associated with supernova explosions in such a population (see discussion below).

\subsection{Physical conditions of the [Fe{\sc\,ii}] and H$_2$ emitting regions}
\label{subsec:physFeH2}

\subsubsection{The origin of the H$_2$ emission}
\label{subsec:physcondH2}

The origin of the  H$_2$ line emission in active galaxies has been investigated in several studies. In most of them it has been concluded that the H$_2$ emission originates in gas excited by thermal processes produced by X-ray and/or shock heating \citep[e.g.][]{hollenbach89,maloney96,ardila04,ardila05,sb09,eso428,n4051,n7582,mrk1066a,mrk1157,mrk79,dors12,rogerio06,rogerio13,mazzalay13a}. Another conclusion of these studies is that H$_2$ emission due to fluorescent excitation \citep{black87} is negligible for most AGNs. 

Here, we have used the H$_2\lambda2.2477\,\mu$m/$\lambda2.1218\,\mu$m line ratio to distinguish between thermal (values between 0.1 and 0.2) and fluorescent ($\approx$0.55) excitation \citep{mouri94,reunanen02,ardila04,ardila05,sb09} in the nuclear region of NGC\,1068.  The map for this ratio is shown in the bottom-right panel of Fig.~\ref{fig:ratios}, presenting values ranging from 0.1 -- typical value at most locations, to 0.45, observed only in a small region surrounding the nucleus.  This map supports thermal excitation for H$_2$ at most locations. At the nucleus, the line ratios are close to those typical of fluorescent excitation, but uncertainties in the H$_2\lambda2.2477\,\mu$m)/$\lambda2.1218\,\mu$m line ratio are larger there, due to the small S/N ratio of the emission-line fluxes. 


The line ratio above can be used to estimate the H$_2$ vibrational temperature $T_{\rm vib}$, while the line ratio  H$_2\lambda2.0338\,\mu$m/$\lambda2.2235\,\mu$m can be used to estimate the rotational temperature $T_{\rm rot}$. These temperatures have similar values for thermal excitation, while for fluorescent excitation, the rotational temperature is smaller than the vibrational temperature, because non-local UV photons overpopulate the highest energy levels compared to the population distribution  expected for a Maxwell-Boltzmann distribution \citep[eg.][]{n4051}. These temperatures can be obtained using
\begin{equation}
T_{\rm rot} \cong -{\rm 1113/ln(0.323\,}\frac{F_{H_{2}\lambda2.0338}}{F_{H_{2}\lambda2.2235}}) 
\end{equation}
and  
\begin{equation}
T_{\rm vib}\cong {\rm 5600/ ln(1.355\,}\frac{F_{H_{2}\lambda2.1218}}{F_{H_{2}\lambda2.2477}}),
\end{equation}
where $F_{\rm H_{2}\lambda2.0338}$, $F_{\rm H_{2}\lambda2.2235}$, $F_{\rm H_{2}\lambda2.1218}$ and $F_{\rm H_{2}\lambda2.2477}$ are the fluxes of the lines \citep{reunanen02}.

\begin{figure*}
    \centering
    \begin{tabular}{cc}
	\includegraphics[scale=0.45]{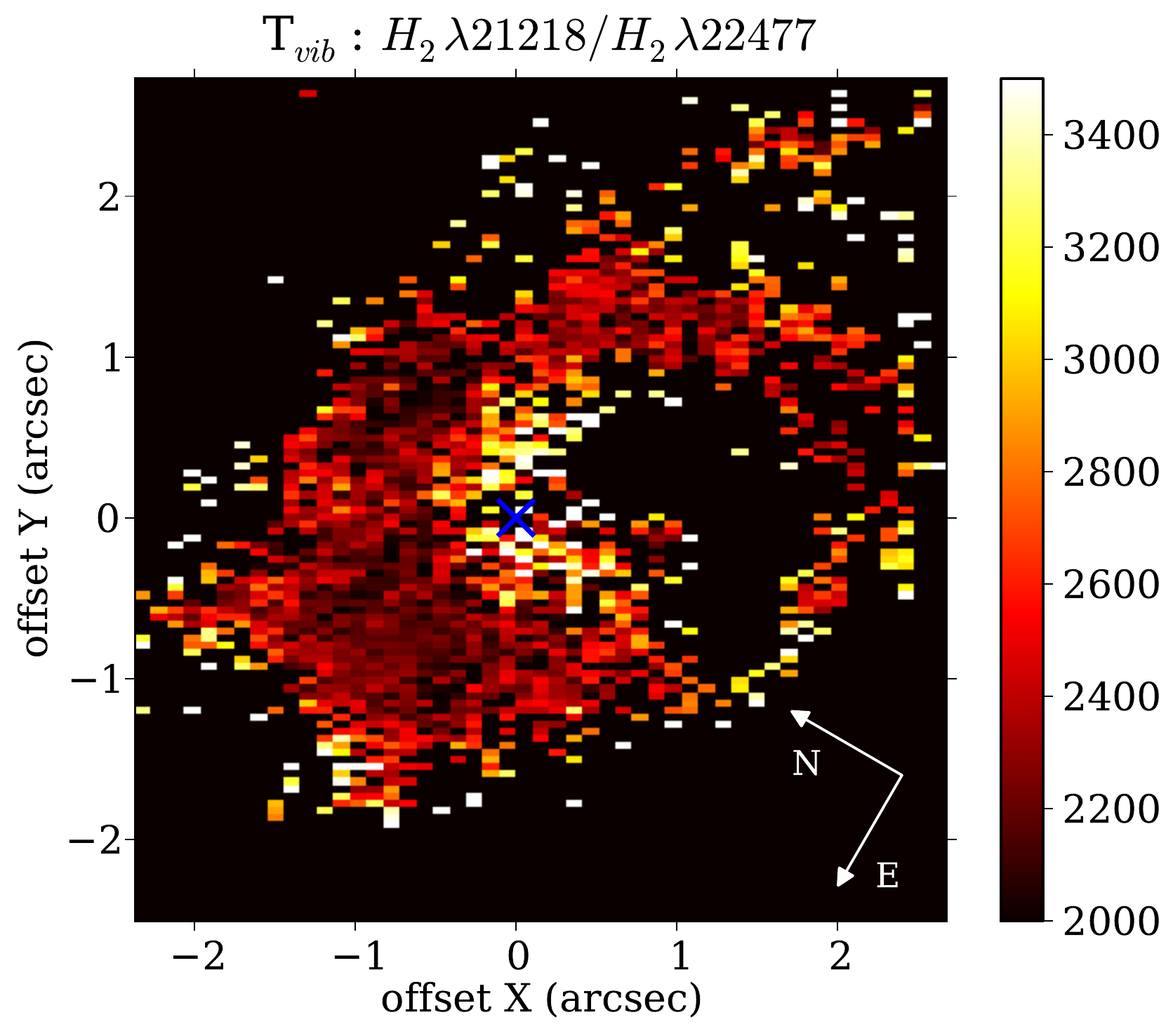}&
	\includegraphics[scale=0.45]{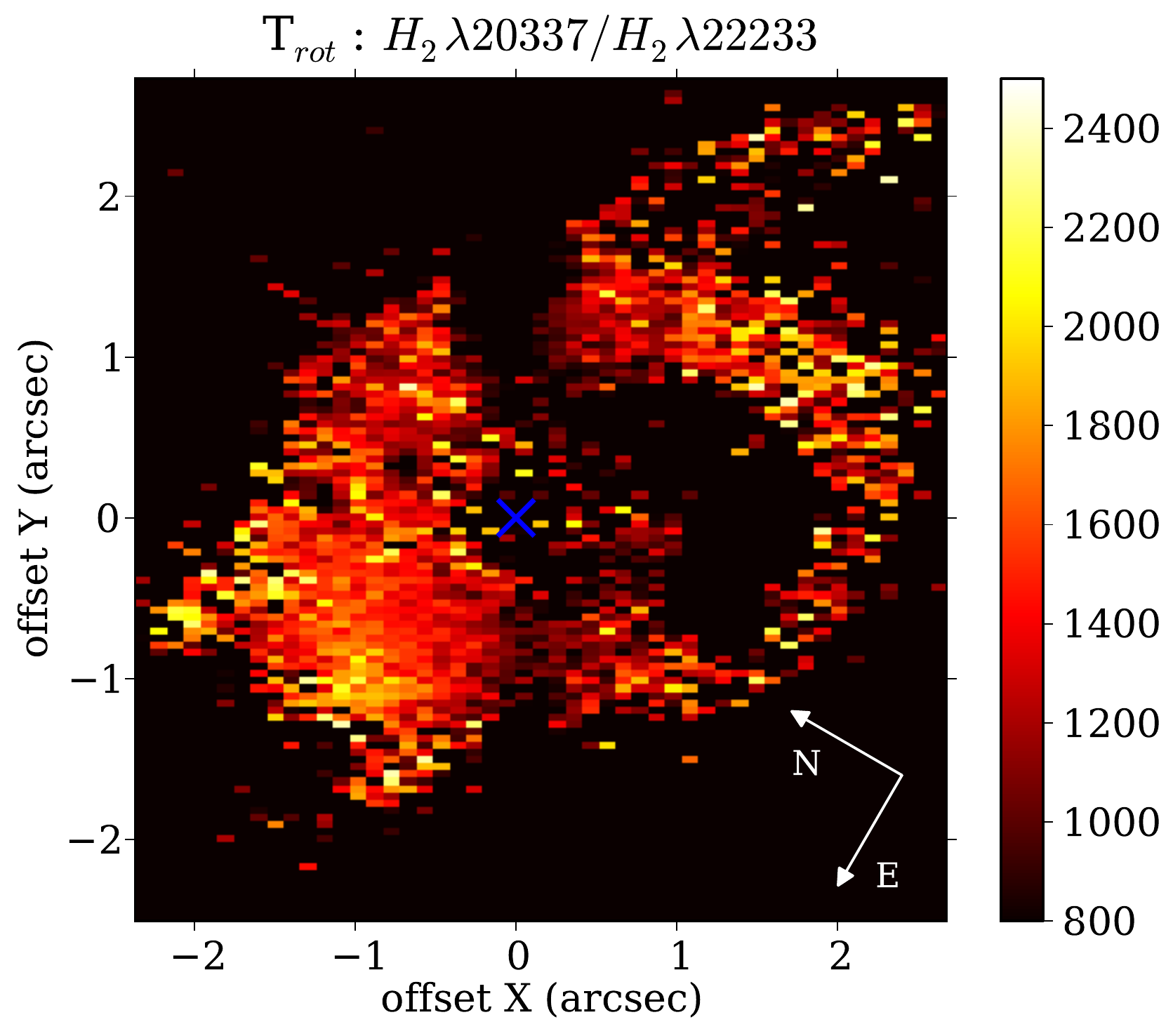}
    \end{tabular}
    \caption{Maps for the vibrational (left panel) and rotational (right panel) temperatures of the H$_2$ emitting gas.}
    \label{fig:trot-vib}
\end{figure*}

In Fig.\,\ref{fig:trot-vib} we present maps for the vibrational (left panel) and rotational (right panel) temperatures obtained using the equations above. The uncertainties in these maps are smaller than 100\,K for most locations. It can be seen that both temperatures are approximately constant over the ring, with somewhat larger values for $T_{\rm vib}$ around the nucleus. This suggests some contribution of fluorescent excitation for the H$_2$ line emission there \citep[e.g.][]{reunanen02,n4051}.

The combined fluxes of the many H$_2$ emission lines that could be measured in our spectra can be used to calculate the thermal excitation temperature using the following equation \citep[e.g.][]{wilman05,sb09,astor}: 

\begin{equation}
 {\rm log}\left(\frac{F_i \lambda_i}{A_i g_i}\right)={\rm constant}-\frac{T_i}{T_{\rm exc}},
\label{eq-ex}
\end{equation}
where $F_i$ is the flux of the $i^{th}$ H$_2$ line, $\lambda_i$ is its wavelength, 
$A_i$ is the spontaneous emission coefficient, $g_i$ is the statistical                         
weight of the upper level of the transition, $T_i$ is the
energy of the level expressed as a temperature and $T_{\rm exc}$ is the excitation temperature. This equation is valid only for thermal excitation, under the assumption of an {\it ortho:para} abundance ratio of 3:1. 

\begin{figure*}
    \begin{tabular}{ccc}
        \includegraphics[scale=0.27]{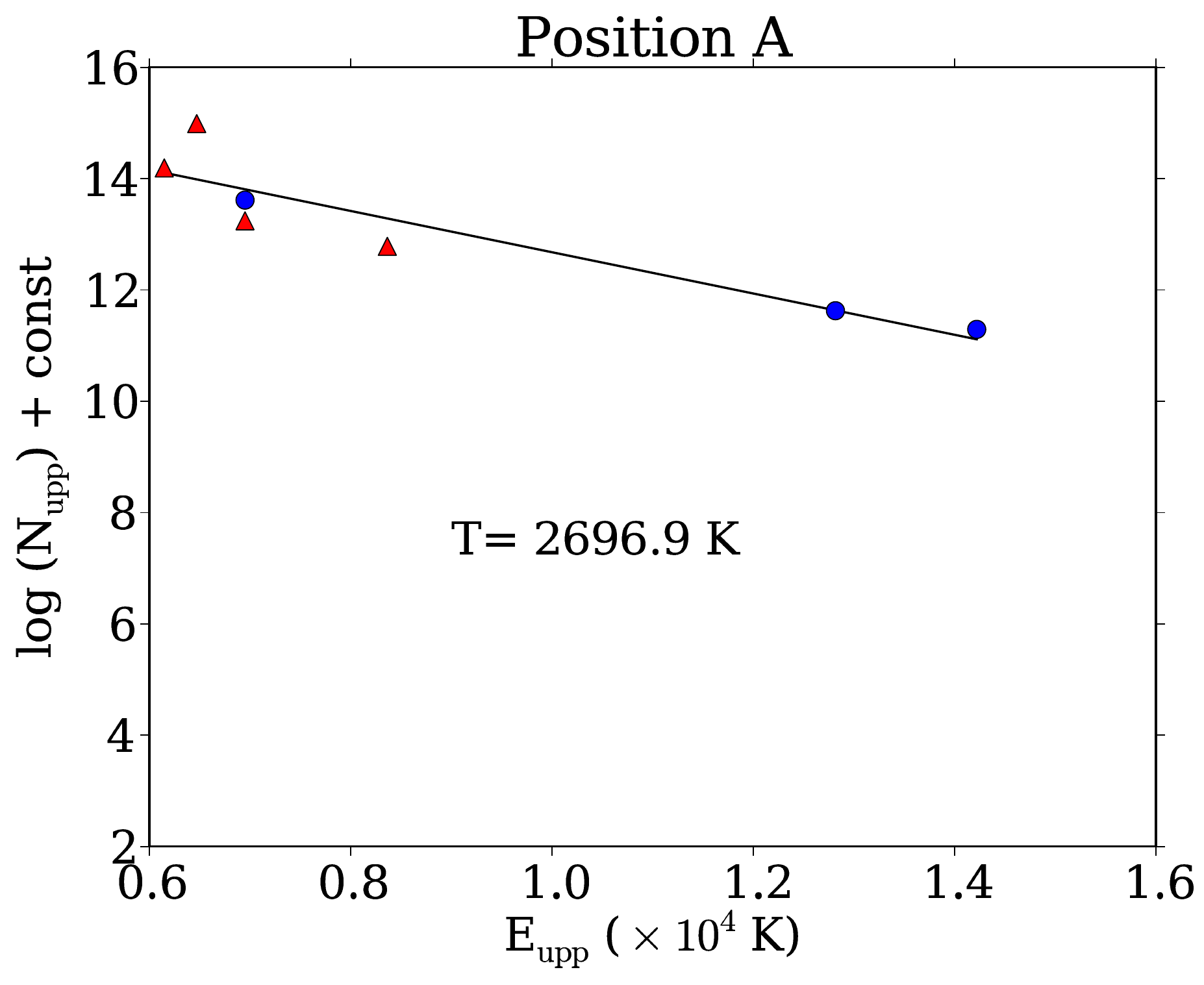}&
        \includegraphics[scale=0.27]{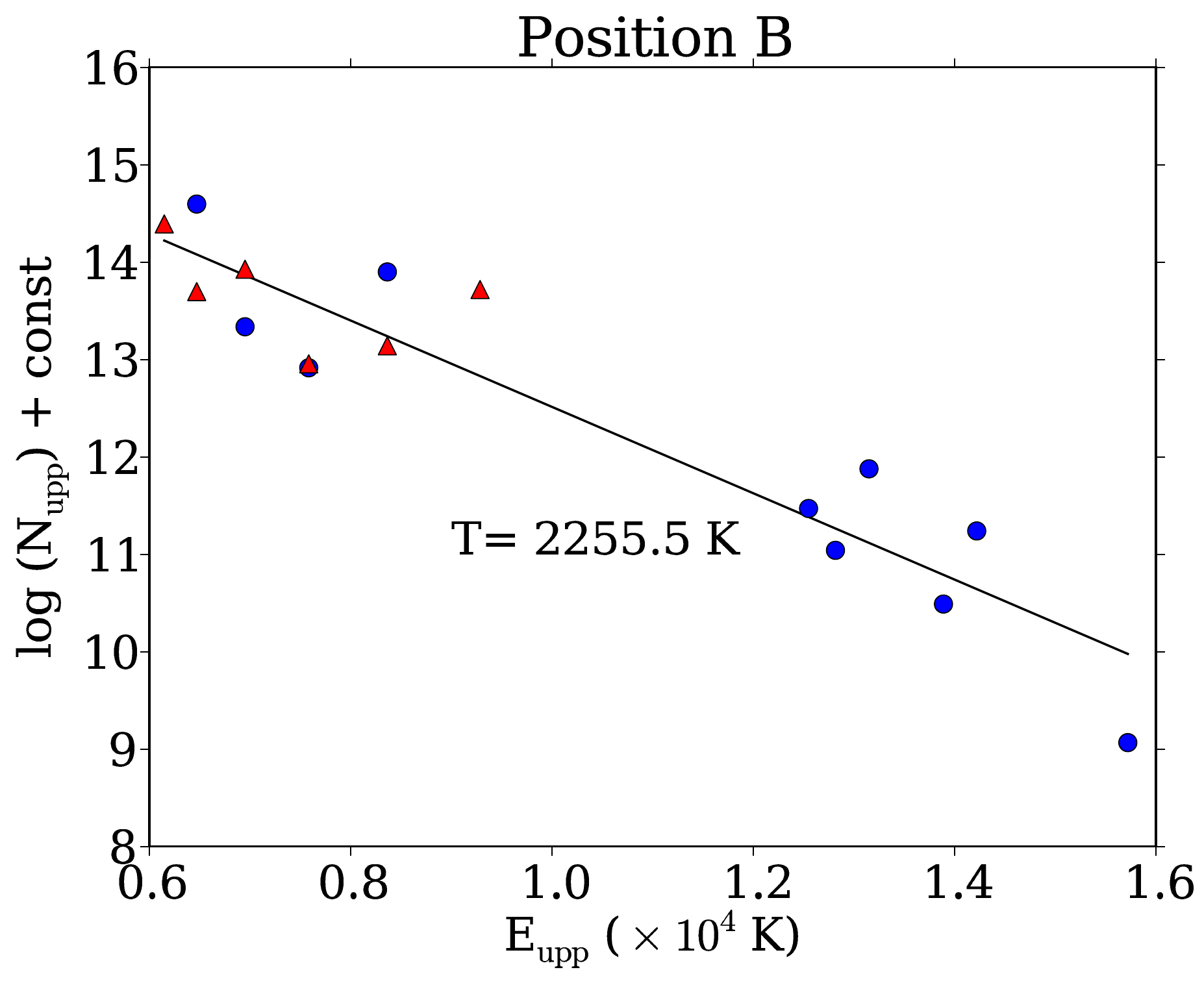}&
        \includegraphics[scale=0.27]{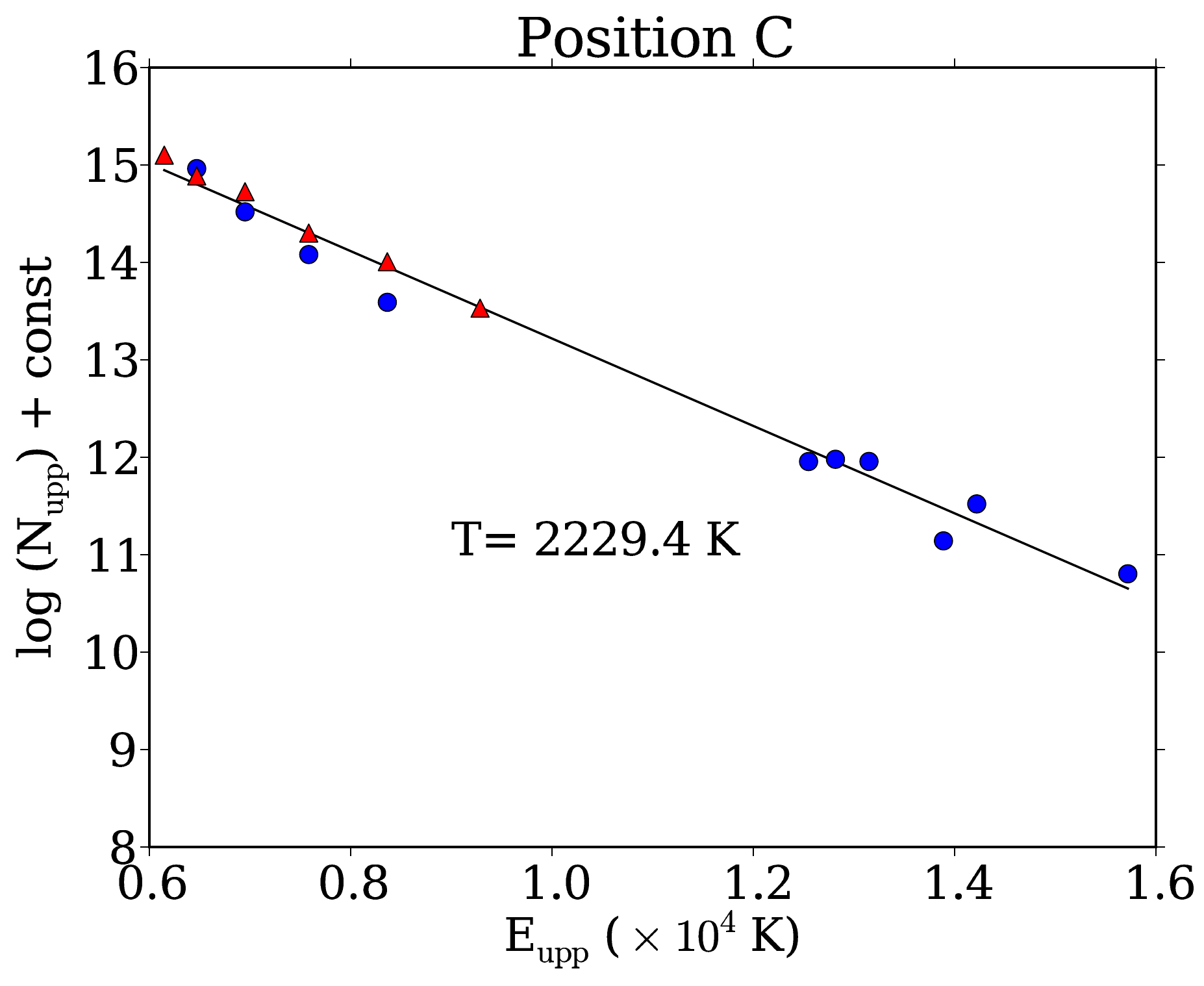}
    \end{tabular}
    \label{fig:temperatures}
    \caption{Thermal excitation temperature derived from the fluxes of a series of H$_2$ emission lines at the three positions shown in Fig. \ref{fig:nuclearzoom}. {\it Ortho} transitions are shown as filled circles and {\it para} transitions
 as  triangles.}
\end{figure*}

In Fig.\,\ref{fig:temperatures} we present a plot of $N_{\rm upp}=\frac{F_i \lambda_i}{A_i g_i}$ (plus an arbitrary constant) $vs$ $E_{\rm upp}={T_i}$ for the locations shown in Fig.~\ref{fig:nuclearzoom}, namely: (1) the nucleus (left panel); (2) the position corresponding to the peak emission of  [Fe\,{\sc ii}] (middle panel); and (3) the location of the peak emission of H$_2$ (right panel).  The fits of the observed data points by the equation~\ref{eq-ex} are shown as continuous lines.  The fit reproduces the observed data for the three positions and thus confirm that the $H_2$ emitting gas is in thermal equilibrium. For the nucleus, the excitation temperature obtained from the fit is $\approx$2700~K, while for the extra-nuclear regions the excitation temperature is about 2250~K. These values for the excitation temperature of H$_2$ in NGC\,1068 are in good agreement with previous estimates of  $T_{\rm exc}$ for other Seyfert galaxies \citep{sb09,mrk1066a,mrk1157,astor}.

In \citet{sb12} we have performed spectral synthesis of the continuum and concluded that, at the ring, there is a strong contribution of a young  but not ionizing, stellar population -- the main component being that with age 30\,Myr. Particularly conspicuous in the H$_2$ flux distribution and in the young population contribution is the ``knot" of strong H$_2$ emission at approximately 1\arcsec\ E of the nucleus. This region shows a large velocity dispersion, being observed from the channel maps in blueshift centered at $-$130\,km\,s$^{-1}$ up to the one in redshift at 112\,km\,s$^{-1}$ and even at 172\,km\,s$^{-1}$. One possibility to explain this high velocity dispersion is the presence of supernova remnants there, what would be compatible with the age of the stellar population. In spite of having a larger velocity dispersion than in the rest of the ring, the H$_2\lambda2.2477\,\mu$m)/$\lambda2.1218\,\mu$m line ratio at this location does not differ from those in the remainder of the ring, supporting also thermal excitation there.

Using long-slit spectroscopy of a sample of 67 emission-line galaxies together with photoionization models,  \citet{rogerio13}  have investigated the origin of the H$_2$ line emission and concluded also that heating by X-rays  is the dominant thermal process for active galaxies. Our more detailed data suggests that this also happens in NGC\,1068:  the  H$_2\lambda2.2477\,\mu$m/$\lambda\,2.1218\,\mu$m ratio also indicates that the H$_2$ emission is mainly due to X-ray heating.

\subsubsection{The origin of the [Fe\,{\sc ii}] emission}

Most of the previous studies referred to in the previous section, address also the excitation mechanisms of the near-IR [Fe\,{\sc ii}] emission lines and in general support a larger contribution of excitation by shocks to the [Fe\,{\sc ii}]  than to the H$_2$ excitation, based mainly on emission-line ratios. However, in \citet{dors12}, we showed that the [Fe\,{\sc ii}] emission in active galaxies can also be reproduced by photoionization models which include only X-rays from the central AGN, even though a contribution from shocks cannot be discarded. On the other hand, \citet{rogerio13} found that shocks play an important role in the origin of the [Fe\,{\sc ii}] emission in AGNs. Thus, the [Fe\,{\sc ii}] origin in AGNs is still an open question to be further investigated.

One interesting property of the [Fe\,{\sc ii}] emission in NGC\,1068 is the fact that, within the FOV of our observations, its emission is more extended than that of [O\,{\sc iii}], as discussed in previous sections. As the Br$\gamma$ flux distribution seems to be similar to that of [O\,{\sc iii}] (see Fig.\,\ref{fig:maps}), we have used the channel maps of Br$\gamma$ \citep{barbosa14} to investigate this difference between the [O\,{\sc iii}] and [Fe\,{\sc ii}] emitting gas also regarding their kinematics. What can be seen in the channel maps is that there is a strong similarity between the two in the blueshift channels, although only at the highest flux levels, as the  emission in the [Fe\,{\sc ii}] map is more extended than that of Br$\gamma$. But the most striking difference is observed in the redshift channels, in which little emission is seen in Br$\gamma$ to the N-NE and S-SW, while in [Fe\,{\sc ii}] there is strong emission within the whole quadrant between N and E. Adopting the approximate geometry for the outflow proposed by \citet{das06}, the origin of this emission in redshift would be the back wall of the conical/houglass outflow to the NE. Actually, the hourglass shape of the outflow seen in the [Fe\,{\sc ii}]  flux distribution has an important contribution from this redshifted part of the outflow.
But why there is emission from this part in [Fe\,{\sc ii}] and not in Br$\gamma$? As pointed out by \citet{mouri00}, the [Fe\,{\sc ii}] emission is excited by electrons in a zone of partially ionized H.  This zone is found around AGN because it is formed when the H ionizing photons have been already absorbed but there is heating of the gas by X-rays and/or shocks, that can still ionize the Fe\,{\sc i}. This explains why we see a lot of [Fe\,{\sc ii}] emission and little Br$\gamma$ emission from the back part of the outflow: most of this region is a partially ionized zone.

Further investigation on the origin of the [Fe\,{\sc ii}] emission can be done via the line ratios shown in Fig.~\ref{fig:ratios}. In particular, the [Fe\,{\sc ii}]$\lambda1.2570\,\mu$m and [P\,{\sc ii}]$\lambda1.1885\,\mu$m emission lines have similar excitation temperatures, with their parent ions having similar ionization potential and radiative recombination coefficients. In H\,{\sc ii} regions, where shocks are not important, the ratio [Fe\,{\sc ii}]/[P\,{\sc ii}] is $\approx\,2$. In supernova remnants, fast shocks destroy the dust grains and release the Fe, increasing its abundance and thus its emission \citep[e.g][]{oliva01,sb09,mrk1157,mrk79}, leading to line ratios larger than 20 \citep{oliva01}. The same can happen in AGNs, where nuclear jets can produce the shocks.
In the top-left panel of Fig.~\ref{fig:ratios}, the ratio [Fe\,{\sc ii}]/[P\,{\sc ii}]$\leq3$ within a circular region with radius 1\arcsec\  centered $\approx$\,0\farcs5 N of the  nucleus, while values of up to 10 are observed further out to the NE and to the SW, along the bipolarl outflow observed in [O\,{\sc iii}] \citep{das06,das07}, suggesting that  shocks are present at these locations.
These shocks are probably due to the interaction of the outflowing material and/or the radio jet with ambient gas. Similar behavior of the [Fe\,{\sc ii}] emission has also been observed for other active galaxies \citep{sb09,mrk1066a,mrk1157,mrk79}: an enhancement of [Fe\,{\sc ii}]/[P\,{\sc ii}] line ratio in association with a radio jet.

The [Fe\,{\sc ii}] (1.2570\,$\mu$m)/Pa$\beta$ line ratio can also be used to investigate the origin of the [Fe\,{\sc ii}] emission.  Values larger than 2 indicate that shocks contribute to the excitation of the [Fe\,{\sc ii}] lines \citep[eg.][]{ardila04,ardila05,eso428,mrk1157,mrk79,sb09}. For NGC\,1068, most regions present values smaller than 2 as can be seen in Fig.~\ref{fig:ratios} indicating that X-rays are the main excitation mechanism of the [Fe\,{\sc ii}] lines. Nevertheless, some higher values are observed to NE along the bicone -- at the same location where the Fe\,{\sc ii}]/[P\,{\sc ii}] presents the highest values -- consistent with shock excitation at these locations.

In summary, we conclude that the [Fe\,{\sc ii}] near-IR lines are originated from the emission of an outflowing gas excited mainly by X-rays from the central AGN, but also with a contribution from shocks associated to the radio jet to NE and SW of the nucleus.

\subsection{Masses of ionized and molecular gas}
\label{subsec:mass}

We can use the fluxes of the Br$\gamma$ and H$_2$ (2.1218$\mu$m) emission lines to derive  the mass of ionized and hot molecular gas in the inner $\approx$400$\times$400~pc$^2$ of NGC\,1068. The mass of ionized gas can be obtained from \citep[e.g.][]{sb09,mrk1066a,mrk79}:

\begin{equation}
 M_{\rm H\,II}=3\times10^{19}\left(\frac{F_{\rm Br\gamma}}{\rm erg\,cm^{-2}\,s^{-1}}\right)\left(\frac{D}{\rm Mpc}\right)^2\left(\frac{N_e}{\rm cm^{-3}}\right)^{-1},
\end{equation}
where $D$ is the distance to the galaxy in cm, $F_{\rm Br\gamma}$ is the Br$\gamma$ flux, units of density are cm$^{-3}$ and we have assumed an electron temperature of 10$^4$K and density in the range $10^2<N_e<10^4$\,cm$^{-3}$. 

Integrating the Br$\gamma$ flux over the NLR, we obtain $F_{\rm Br\gamma}\approx1.6\times10^{-15}\,{\rm erg s^{-1} cm^{-2}}$ resulting in $M_{\rm H\,II}\approx2.2\times10^{4} {\rm M_\odot}$, for an electron density $N_e=500\,{\rm cm^{-3}}$ -- the average $N_e$ we have found in our previous studies \citep{sb09,allan11,allan14a,allan14b}.

The mass of hot molecular gas, that emits the near-IR H$_2$ lines, can be obtained from \citep[e.g.][]{scoville82,sb09,n4051,mrk79}:
 
\begin{equation} 
M_{H_2}=5.0776\times10^{13}\left(\frac{F_{H_{2}\lambda2.1218}}{\rm erg\,s^{-1}\,cm^{-2}}\right)\left(\frac{D}{\rm Mpc}\right)^2,
\label{mh2}
\end{equation}
where $F_{H_{2}\lambda2.1218}$ is the H$_2$ (2.1218$\mu$m) emission-line flux, $D$ is the distance to the galaxy and M$_{H_2}$ is given in solar masses. 

The integrated flux of the H$_2$ line is $F_{H_{2}\lambda2.1218}\approx2.5\times10^{-15}\,{\rm erg s^{-1} cm^{-2}}$ for NGC\,1068, resulting in a mass of only $M_{H_2}\approx29\,{\rm M_\odot}$. 

Thus, the mass in ionized gas is about 1000 times larger than the mass in hot molecular gas in the inner 200\,pc of NGC\,1068. The ratio $ M_{\rm H\,II}/M_{\rm H_2}$ is similar to those we have found in similar studies of other active galaxies, which ranges from $10^3$ to 10$^4$ \citep{sb09,mrk1066a,mrk79}.

As pointed out in our previous studies, the H$_2$ emission observed in the near-IR, originates from the "hot skin" of a much larger mass reservoir dominated by cold molecular gas. Previous studies of many active galaxies, in which masses of both cold and hot molecular gas have been obtained, imply factors of 10$^5$ to 10$^7$ between the mass of cold and hot molecular gas \citep[e.g.][]{dale05,mazzalay13a}. Using the expression proposed by \citet{mazzalay13a}, that already take into account this factor, we can obtain the mass of cold molecular gas in the inner 200\,pc of NGC\,1068:

\begin{equation}
M_{\rm cold} \approx1174\left(\frac{L_{H_{2}\lambda2.1218}}{\rm L_\odot}\right),
\label{mh2}
\end{equation}
where $L_{H_{2}\lambda2.1218}$ is the luminosity of the H$_2$ line and the mass is given in solar masses. 

Using the flux quoted above for the H$_2\lambda2.1218\mu$m line, we obtain $L_{H_{2}\lambda2.1218}\approx6.9\times37\,{\rm erg\, s^{-1}}=1.8\times10^{4}\,{\rm L_\odot}$ and thus $M_{\rm cold}\approx2.1\times10^{7}\,{\rm M_\odot}$. This value is $10^6$ times larger than that obtained for the mass of hot molecular gas,  in agreement with the range quoted above (between 10$^5$ to 10$^7$) \citep[e.g.][]{dale05,mazzalay13a,mrk79}.

\section{Final Remarks}
\label{sec:conclusions}

We have mapped the emitting gas flux distributions, reddening and excitation of the inner $\approx200$~pc  of NGC\,1068, at a spatial resolution of $\approx$10\,pc using near-IR adaptive optics integral field spectroscopy obtained at Gemini with the instrument NIFS.  The main conclusions of this paper are:

\begin{itemize}
\item The emitting H$_2$ and ionized gases have completely different flux distributions. The first is observed in a circumnuclear ring in the plane of the galaxy with radius 75\,pc and the latter trace the bipolar outflows previously observed in the optical;

\item The comparison of line-ratio maps with the values predicted by different excitation models shows that most of the H$_2$ line emission is excited by thermal processes, mainly due to heating of the gas by X-rays from the central AGN to an excitation temperature of 2200\,K;

\item The flux distribution in the H emission lines, such as Br$\gamma$, follows that observed in the optical [O\,{\sc iii}]  emission, previously described as having  a bipolar cone-shaped structure; the [Fe\,{\sc ii}] flux distribution is nevertheless more extended and shows an hourglass structure, broader than a cone and similar to that seen in planetary nebulae such as NGC\,6302;

\item  The broader and more extended flux distribution in the [Fe\,{\sc ii}] emission lines is attributed to the fact that the it originates in a partially ionized region, that extends beyond the fully ionized region and is excited mainly by X-rays from the central AGN, with some contribution from shocks to NE and SW of the nucleus along the bicone axis. Shock excitation is supported by the enhancement in the [Fe\,{\sc ii}](1.2570\,$\mu$m)/[P\,{\sc ii}](1.1885\,$\mu$m) and [Fe\,{\sc ii}](1.2570\,$\mu$m)/Pa$\beta$ emission-line ratios, and is attributed to the passage of the radio jet through the NLR;

\item  Reddening maps for the NLR were obtained from H and [Fe\,{\sc ii}] emission-line ratios, presenting values in the range $0\le\,E(B-V)\,\le\,2$, but with an average value of 0.6, observed at most locations; the highest reddening values are observed at $\approx$\,0\farcs8 north of the nucleus; 

\item The mass of of ionized gas in the inner $\approx200\times200$~pc$^2$ of NGC\,1068 is  $M_{\rm H\,II}\approx2.2\times10^{4} {\rm M_\odot}$, while the mass of the H$_2$ emitting gas (hot molecular gas) is only  $M_{H_2}\approx29\,{\rm M_\odot}$. Considering the average factor  between the masses of cold and hot molecular gas observed for AGN in general, the total (dominated by the cold) H$_2$ mass is estimated to be $M\approx2\times10^{7}\,{\rm M_\odot}$.

\end{itemize}

\section*{Acknowledgments}
Based on observations obtained at the Gemini Observatory, which is
operated by the Association of Universities for Research in Astronomy,
Inc., under a cooperative agreement with the NSF on behalf of the
Gemini partnership: the National Science Foundation (United States),
the Science and Technology Facilities Council (United Kingdom), the
National Research Council (Canada), CONICYT (Chile), the Australian
Research Council (Australia),  Minist\'erio da Ci\^encia e Tecnologia (Brazil) and SECYT
(Argentina). This work has been partially supported by the Brazilian
institutions CNPq and FAPERGS.

{}   

\begin{thebibliography}{}

\bibitem[\protect\citeauthoryear{Alloin et al.}{2001}]{alloin01} Alloin, D., Galliano, E., Cuby, J. G., Marco, O.,  Rouan, D., Cl\'enet, Y., Granato, G. L. \& Franceschini, A., 2001, A\&A, 369, L33.

\bibitem[\protect\citeauthoryear{Antonucci \& Miller}{1985}]{antonucci1985} Antonnucci, R. R. J., Miller, J. S., 1985, ApJ, 297, 621

\bibitem[\protect\citeauthoryear{Antonucci}{1993}]{antonucci1993} Antonnucci, R. R. J., 1993, Annu. Rev. Astron. Astrophys., 31, 473

\bibitem[\protect\citeauthoryear{Axon et al.}{1998}]{axon98} Axon, D. J., Marconi, A., Capetti, A., Macchetto, F., Schreier, E., \& Robinson, A. 1998, ApJ, 496, L75

\bibitem[\protect\citeauthoryear{Bautista \& Pradhan}{1998}]{bautista98} Bautista, M. A.  \& Pradhan, A. K. ,1998, ApJ, 492, 650

\bibitem[\protect\citeauthoryear{Barbosa et al.}{2014}]{barbosa14} Barbosa, F., Storchi-Bergmann, T., McGregor, P., 2014, MNRAS, submitted.

\bibitem[\protect\citeauthoryear{Black \& van Dishoeck}{1987}]{black87} Black, J. H., \& van Dishoeck, E. F.  1987, ApJ, 322, 412

\bibitem[\protect\citeauthoryear{Capetti et al.}{1997}]{capetti1997} Capetti, A., Macchetto, F. D., Lattanzi, M. G., 1997, Ap\&SS, 248, 245

\bibitem[\protect\citeauthoryear{Cardelli, Clayton \& Mathis}{1989}]{cardelli89} Cardelli, J. A., Clayton, G. C. \& Mathis, J. S., 1989, ApJ, 345, 245

\bibitem[\protect\citeauthoryear{Cecil et al.}{2002}]{cecil02} Cecil, G., Dopita, M. A., Groves, B., Wilson, A. S., Ferruit, P., P\'econtal,
E., \& Binette, L. 2002, ApJ, 568, 627.

\bibitem[\protect\citeauthoryear{Dale et al.}{2005}]{dale05} Dale, D. A., Sheth, K., Helou, G., Regan, M. W., \& H\"uttemeister, S., 2005, ApJ, 129, 2197

\bibitem[\protect\citeauthoryear{Das et al.}{2006}]{das06} Das, V., Crenshaw, D. M., Kraemer, S. B., Deo, R. P., 2006, AJ, 132, 620

\bibitem[\protect\citeauthoryear{Das et al.}{2007}]{das07} Das, V., Crenshaw, D. M., Kraemer, 2007, ApJ, 656, 699


\bibitem[\protect\citeauthoryear{de Vaucouleurs et al.}{1991}]{devaucouleurs1991} de Vaucouleurs et al. 1991, Third Reference Catalogue of Bright Galaxies. (Volume II), ISBN 0-387-97550-0

\bibitem[\protect\citeauthoryear{Dors et al.}{2012}]{dors12} Dors, O. L., Riffel, Rogemar A., Cardaci, M. C., H\"agele, G. F., Krabbe, A. C., P\'erez-Montero, E. \& Rodrigues, I., 2012, MNRAS, 422, 252.


\bibitem[\protect\citeauthoryear{Emsellem et al.}{2006}]{emsellem06}  Emsellem, E., Fathi, K., Wozniak, H., Ferruit, P., Mundell, C. G., Schinnerer, E. 2006, 
MNRAS, 365, 367.

\bibitem[\protect\citeauthoryear{Evans et al.}{1991}]{evans91} Evans, I. N., Ford, H. C., Kinney, A. L., Antonucci, R. R. J., Armus, L., \& Gaganoff, S., 1991, ApJ, 369, L27.

\bibitem[\protect\citeauthoryear{Evans et al.}{1993}]{evans93} Evans, I. N. et al., 1993, ApJ, 417, 82

\bibitem[\protect\citeauthoryear{Fathi et al.}{2006}]{fathi06} Fathi, K., Storchi-Bergmann, T., Riffel, R. A., Winge, C., Axon, D. J.,  Robinson, A., Capetti, A., \& Marconi, A., 2006, ApJL, 641, L25.

\bibitem[\protect\citeauthoryear{Gerssen et al.}{2006}]{gerssen06}  Gerssen, J., Allington-Smith, J., Miller, B. W., Turner, J. E. H., Walker, A., 2006, MNRAS, 365, 29.

\bibitem[\protect\citeauthoryear{Galliano \& Alloin}{2002}]{galliano02} Galliano, E. \& Alloin, D., 2002, A\&A, 393, 43.

\bibitem[\protect\citeauthoryear{Gallimore et al.}{1996}]{gallimore96} Gallimore, J. F., Baum, S. A., O'Dea, C. P., \& Pedlar, A. 1996, ApJ, 458, 136.

\bibitem[\protect\citeauthoryear{Gallimore et al.}{2004}]{gallimore2004} Gallimore, J. F., Baum, S. A., O'Dea, C. P., 2004, ApJ, 613, 794

\bibitem[\protect\citeauthoryear{Ho, Filippenko \& Sargent}{1997}]{ho97} Ho, L. C., Filippenko, A. V., Sargent, W. L., 1997, ApJS, 112, 31.

\bibitem[\protect\citeauthoryear{Hollenbach \& McKee}{1989}]{hollenbach89} Hollenbach, D., \& McKee, C. F., 1989, ApJ, 342, 306.

\bibitem[\protect\citeauthoryear{Kormendy, Bender \& Cornell}{2011}]{kormendy11} Kormendy, J., Bender, R., Cornell, M. E., 2011, Nature, 469, 374.

\bibitem[\protect\citeauthoryear{Kraemer \& Crenshaw}{2000}]{kraemer2000} Kraemer, S. B. , Crenshaw, D. M., 2000, ApJ, 532, 256

\bibitem[\protect\citeauthoryear{Krips et al.}{2011}]{krips11} Krips, M. et al., 2011, ApJ, 736, 37.

\bibitem[\protect\citeauthoryear{Koski}{1978}]{koski78} Koski, A. T. 1978, ApJ, 223, 56.

\bibitem[\protect\citeauthoryear{Lodato \& Bertin}{2003}]{lodato2003} Lodato, G., Bertin, G., A\&A, 398, p.517

\bibitem[\protect\citeauthoryear{Macchetto et al.}{1994}]{macchetto94} Macchetto, F., Capetti, A., Sparks, W. B., Axon, D. J., \& Boksenberg, A., 1994, ApJ, 435, L15.

\bibitem[\protect\citeauthoryear{Maloney, Hollenbach \& Tielens}{1996}]{maloney96} Maloney, P. R.,  Hollenbach, D. J.,  \& Tielens, A. G. G. M., 1996, ApJ, 466, 561.


\bibitem[\protect\citeauthoryear{Martins et al.}{2010}]{martins10} Martins, L. P., Rodr\'iguez-Ardila, A., de Souza, R., Gruenwald, R., 2010, MNRAS, 406, 2168.

\bibitem[\protect\citeauthoryear{Mazzalay et al.}{2013a}]{mazzalay13a} Mazzalay, X. et al., 2013a, MNRAS, 428, 2389

\bibitem[\protect\citeauthoryear{Mazzalay et al.}{2013b}]{mazzalay13b} Mazzalay, X. Rodr\'iguez-Ardila, A., Komossa, S. \& McGregor, P. J., 2013b, MNRAS, 430, 2411.

\bibitem[\protect\citeauthoryear{McGregor et al.}{2003}]{nifs03}  McGregor, P. J. et al., SPIE, 2003, 4841, 1581, eds. Iye, M. \& Moorwood, A. F. M.

\bibitem[\protect\citeauthoryear{Mouri}{1994}]{mouri94} Mouri, H., 1994, ApJ, 427, 777.

\bibitem[\protect\citeauthoryear{Mouri, Kawara \& Taniguchi}{2000}]{mouri00} Mouri, H., Kawara, K., \& Taniguchi, Y.  2000, ApJ, 528, 186.

\bibitem[\protect\citeauthoryear{M\"uller S\'anchez et al.}{2009}]{ms09} M\"uller S\'anchez, F., Davies, R. I., Genzel, R., Tacconi, L. J., Eisenhauer, F., Hicks, E. K. S., Friedrich, S., 
\& Sternberg, A., 2009, ApJ, 691, 749.

\bibitem[\protect\citeauthoryear{Muxlow et al.}{1996}]{muxlow96} Muxlow, T. W. B., Pedlar, A., Holloway, A. J., Gallimore, J. F., \& Antonucci, R. R. J. 1996, MNRAS, 278, 854


\bibitem[\protect\citeauthoryear{Nussbaumer \& Storey}{1988}]{nussbaumer88}  Nussbaumer, H. \& Storey, P. J. 1988, A\&A, 193, 327

\bibitem[\protect\citeauthoryear{Oliva et al.}{2001}]{oliva01} Oliva, E. et al., 2001, A\&A, 369, L5. 

\bibitem[\protect\citeauthoryear{Osterbrock \& Ferland}{2006}]{osterbrock06} Osterbrock, D. E. \& Ferland, G. J., 2006, 
Astrophysics of Gaseous Nebulae and Active Galactic Nuclei, Second Edition, University Science Books, Mill Valley, California.

\bibitem[\protect\citeauthoryear{Pogge}{1988}]{pogge88} Pogge, R. W., 1988, ApJ, 328, 519.

\bibitem[\protect\citeauthoryear{Reunanen, Kotilainen \& Prieto}{2002}]{reunanen02} Reunanen, J., Kotilainen, J. K., \& Prieto, M., A., 2002, MNRAS, 331, 154


\bibitem[\protect\citeauthoryear{Riffel et al.}{2006}]{eso428} Riffel, Rogemar A., Sorchi-Bergmann, T., Winge, C., Barbosa, F. K. B., 2006, MNRAS, 373, 2.

\bibitem[\protect\citeauthoryear{Riffel et al.}{2008}]{n4051} Riffel, R. A., Storchi-Bergmann, T., Winge, C., McGregor, P., Beck, T. \& Schmitt, H., 2008, MNRAS, 385, 1129

\bibitem[\protect\citeauthoryear{Riffel et al.}{2009}]{n7582} Riffel, Rogemar A., Storchi-Bergmann, T., Dors, O. L., Winge, C., 2009, MNRAS, 393, 783.

\bibitem[\protect\citeauthoryear{Riffel, Storchi-Bergmann \& Nagar}{2010a}]{mrk1066a} Riffel, Rogemar A., Storchi-Bergmann, T. \& Nagar, N. M., 2010, MNRAS, 404, 166.

\bibitem[\protect\citeauthoryear{Riffel et al.}{2010b}]{mrk1066pop} Riffel, Rogemar A. \& Storchi-Bergmann, T., Riffel, R., \& Pastoriza, M. G., 2010, ApJ, 713, 469.

\bibitem[\protect\citeauthoryear{Riffel \& Storchi-Bergmann}{2011a}]{mrk1066b} Riffel, Rogemar A. \& Storchi-Bergmann, T., 2011, MNRAS, 411, 469.


\bibitem[\protect\citeauthoryear{Riffel \& Storchi-Bergmann}{2011b}]{mrk1157} Riffel, Rogemar A. \& Storchi-Bergmann, T., 2011, MNRAS, 417, 2752.

\bibitem[\protect\citeauthoryear{Riffel, Storchi-Bergmann \& Winge}{2013}]{mrk79} Riffel, R. A., Storchi-Bergmann, T., Winge, C., 2013, 430, 2249.

\bibitem[\protect\citeauthoryear{Riffel, Storchi-Bergmann \& Riffel}{2013}]{n5929} Riffel, R. A., Storchi-Bergmann, T., Riffel, R., 2014, ApJ, 780L, 24. 


\bibitem[\protect\citeauthoryear{Riffel, Rodr\'\i guez-Ardila, \& Pastoriza}{2006}]{rogerio06} Riffel, R., Rodr\'\i guez-Ardila, A. \& Pastoriza, M. G., 2006, A\&A, 457, 61.

\bibitem[\protect\citeauthoryear{Riffel et al.}{2011c}]{mrk1157b} Riffel, R., Riffel, Rogemar A., Ferrari, F., \& Storchi-Bergmann, T., 2011, MNRAS, 416, 493.


\bibitem[\protect\citeauthoryear{Riffel et al.}{2013}]{rogerio13} Riffel, R., Rodr\'iguez-Ardila, A., Aleman, I., Brotherton, M. S., Pastoriza, M. G., Bonatto, C., Dors, O. L., 2013, MNRAS, 432, 3545.


\bibitem[\protect\citeauthoryear{Rodr\'\i guez-Ardila et al.}{2004}]{ardila04} Rodr\'\i guez-Ardila, A.,  Pastoriza, M. G., Viegas, S., Sigut, T. A. A., \& Pradhan, A. K., 2004,  A\&A, 425, 457.

\bibitem[\protect\citeauthoryear{Rodr\'\i guez-Ardila, Riffel \& Pastoriza}{2005}]{ardila05} Rodr\'\i guez-Ardila, A., Riffel, R., \& Pastoriza, M. G. 2005,  MNRAS, 364, 1041.


\bibitem[\protect\citeauthoryear{Schinnerer et al.}{2000}]{schinnerer00} Schinnerer, E., Eckart, A., Tacconi, L. J., Genzel, R. \& Downes, D., 2000, ApJ, 533, 850.

\bibitem[\protect\citeauthoryear{Schmitt et al.}{2003}]{sda2003} Schmitt, H. R., Donley, J. L., Antonnucci, R. R. J., Hutchings, J. B.; Kinney, A. L., 2003, ApJS, 148, 327

\bibitem[\protect\citeauthoryear{Sch\"onel J\'unior et al}{2013}]{astor} Schonel J\'unior, A. J., Riffel, R. A., Stochi-Bergmann, T., Winge, C., 2013, MNRAS,  submitted

\bibitem[\protect\citeauthoryear{Schnorr-M\"uller et al}{2011}]{allan11} Schnorr-M\"uller, A., Storchi-Bergmann, T., Riffel, R. A., Ferrari, F., Steiner, J. E., Axon, D. J., Robinson, A., 2010, MNRAS, 413, 149.

\bibitem[\protect\citeauthoryear{Schnorr-M\"uller et al}{2014a}]{allan14a} Schnorr-M\"uller, A., Storchi-Bergmann, T., Nagar, N. M., Robinson, A., Lena, D., Riffel, R. A., Couto, G. S., 2014a, MNRAS, 437, 1708.

\bibitem[\protect\citeauthoryear{Schnorr-M\"uller et al}{2014b}]{allan14b} Schnorr-M\"uller, A., Storchi-Bergmann, T., Nagar, N. M., Ferrari, F., 2014b, MNRAS, 438, 3322.


\bibitem[\protect\citeauthoryear{Scoville et al.}{1982}]{scoville82} Scoville, N. Z., Hall, D. N. B., Kleinmann, S. G., \& Ridgway, S. T. 1982, 253, 136

\bibitem[\protect\citeauthoryear{Simpson et al.}{2002}]{simpson02} Simpson, J. P., Colgan, S. W. J., Erickson, E. F., Hines, D. C., Schultz, A. S. B., Trammell, S. R., 2002, ApJ, 574, 95.

\bibitem[\protect\citeauthoryear{Storchi-Bergmann et al.}{2007}]{sb07} Storchi-Bergmann, T., Dors Jr., O.,  Riffel,  R. A., Fathi, K.,  Axon, D. J., \& Robinson, A., 2007, ApJ, 670,  959.

\bibitem[\protect\citeauthoryear{Storchi-Bergmann et al.}{2009}]{sb09} Storchi-Bergmann, T., McGregor, P. J.,  Riffel,  R. A., Sim\~ oes Lopes, R., Beck, T., \& Dopita, M., 2009, MNRAS, 394, 1148

\bibitem[\protect\citeauthoryear{Storchi-Bergmann et al.}{2010}]{sb10} Storchi-Bergmann, T., Sim\~oes Lopes, R., McGregor, P. Riffel, Rogemar A.,
Beck, T., Martini, P., 2010, MNRAS, 402, 819.

\bibitem[\protect\citeauthoryear{Storchi-Bergmann et al.}{2012}]{sb12}  Storchi-Bergmann, T., Riffel, Rogemar A., Riffel, R. ; Diniz, M. R. ; Vale, T. B., McGregor, P. J. 2012, ApJ, 755, 87.

\bibitem[\protect\citeauthoryear{Young et al.}{2001}]{young01} Young, A. J., Wilson, A. S., \& Shopbell, P. L. 2001, ApJ, 556, 6.

\bibitem[\protect\citeauthoryear{Wilman, Edge \& Johnstone}{2005}]{wilman05} Wilman, R. J., Edge, A. C. \& Johnstone, R. M 2005, MNRAS, 359, 755

\bibitem[\protect\citeauthoryear{Worden et al.}{1984}]{worden84} Worden, E.F., Comaskey, B., Densberger, J., Christensen, J., McAfee, J.M., Paisner, J.A., Conway, J.G. 1984, J. Opt. Soc. Am. B: Opt. Phys., 1:2, 314.
   
\end{thebibliography}
\end{document}